\documentclass[sigconf]{acmart}
\AtBeginDocument{%
	}

\renewcommand\footnotetextcopyrightpermission[1]{}
\usepackage{
	algorithm,
	algorithmic,
	amsmath,
    amsthm,
	array,
	balance,
	bm,
	booktabs,
	color,
	comment,
	fancyhdr,
	filecontents,
	float,
	framed,
	enumerate,
	enumitem,
	epsfig,
	wrapfig,
	graphics,
	graphicx,
	hyperref,
	ifthen,
	latexsym,
	makecell,
	multirow,
	setspace,
	subcaption,
	textcomp,
	tikz,
	url,
	xparse,
	xspace
}
\newcommand{\secref}[1]{\mbox{Section~\ref{#1}}}

\newcommand{\figref}[1]{\mbox{Figure~\ref{#1}}}
\renewcommand{\eqref}[1]{\mbox{Equation~\ref{#1}}}

\newcommand{\tabref}[1]{\mbox{Table~\ref{#1}}}

\DeclareMathOperator*{\argmin}{argmin}
\DeclareMathOperator*{\argmax}{argmax}
\expandafter\def\csname ver@subfig.sty\endcsname{}
\usepackage{svg}

\usepackage{titlesec}

\newcommand\guozhu[1]{\textcolor{blue}{GZ: #1}}
\newcommand\xff[1]{\textcolor{green}{XF: #1}}

\newcommand{\tool}{\textsc{AuthNet}\xspace}
\newcommand{\sys}{\textsc{AuthNet}\xspace}

\newtheorem{mydef}{Definition}
\theoremstyle{fancyindent}
\newtheorem{mypro}{Property}

\begin{document}

\title{\tool: Neural Network with Integrated Authentication Logic}

\author{Yuling Cai}
\affiliation{%
  \institution{Institute of Information Engineering, Chinese Academy of Sciences}
  \city{Beijing}
  \country{China}
}
\email{caiyuling@iie.ac.cn}

\author{Fan Xiang}
\affiliation{%
  \institution{Institute of Information Engineering, Chinese Academy of Sciences}
  \city{Beijing}
  \country{China}
}
\email{xiangfan@iie.ac.cn}

\author{Guozhu Meng}
\affiliation{%
  \institution{Institute of Information Engineering, Chinese Academy of Sciences}
  \city{Beijing}
  \country{China}
}
\email{mengguozhu@iie.ac.cn}

\author{Yinzhi Cao}
\affiliation{%
  \institution{Johns Hopkins University}
  \city{Baltimore}
  \state{Maryland}
  \country{USA}
}
\email{yinzhi.cao@jhu.edu}

\author{Kai Chen}
\affiliation{%
  \institution{Institute of Information Engineering, Chinese Academy of Sciences}
  \city{Beijing}
  \country{China}
}
\email{chenkai@iie.ac.cn}

\begin{abstract}


Model stealing, i.e., unauthorized access and exfiltration of deep learning models, has become one of the major threats. 
Proprietary models may be protected by access controls and encryption. 
However, in reality, these measures can be compromised due to system breaches, query-based model extraction or a disgruntled insider. 
Security hardening of neural networks is also suffering from limits, for example, model watermarking is passive, cannot prevent the occurrence of piracy and not robust against transformations.

To this end, we propose a native authentication mechanism, called \sys, which integrates authentication logic as part of the model without any additional structures. 
Our key insight is to reuse redundant neurons with low activation and embed authentication bits in an intermediate layer, called a gate layer.  Then, \sys fine-tunes the layers after the gate layer to embed authentication logic so that only inputs with special secret key can trigger the correct logic of \sys. 
It exhibits two intuitive advantages. It provides the last line of defense, i.e., even being exfiltrated, the model is not usable as the adversary cannot generate valid inputs without the key. 
Moreover, the authentication logic is difficult to inspect and identify given millions or billions of neurons in the model.
We theoretically demonstrate the high sensitivity of \tool to the secret key and its high confusion for unauthorized samples. \tool is compatible with any convolutional neural network, where our extensive evaluations show that \sys successfully achieves the goal in rejecting unauthenticated users (whose average accuracy drops to 22.03\%) with a trivial accuracy decrease (1.18\% on average) for legitimate users, and is robust against model transformation and adaptive attacks.

\end{abstract}

\maketitle
\fancyhf{}

\renewcommand{\acmConference}{}{}

\section{Introduction}
\label{sec:intro}


Deep learning has gained tremendous success in image and video processing~\cite{9609683,yang2019cdeeparch,he2018amc}, face detection~\cite{Yang_2015_ICCV,Li_2015_CVPR,smith2015face,soliman2013face}, speech recognition~\cite{10.1145/3178115,6639344,ai2019speaker,lei2013accurate}, and so on~\cite{8003957,Wang2019LearningDT,tan2022dynamic,zhao2022survey}. 
To be accessible for end users, models are deployed into a variety of cloud platforms (e.g., AWS, Google Cloud, Microsoft Azure) and on-premises devices (e.g., mobile phones, smart home, electrical equipment).
Cloud-based deployment may provide Machine Learning as a Service (MLaaS) for thousands of internet users who can use the service with an authorized API key. On-device models are usually required for privacy preservation, low latency and unstable network connectivity.

%
%
However, deep learning models are under the significant threats after deployment due to model theft and misuse ~\cite{8953839,prada2019,usenix2021drmi,9519386,10.1145/3466752.3480112,pmlr-v162-lee22e,sun2023shadownet} that could cause devastating consequences including monetary loss and privacy leakage for model owners ~\cite{survey2022,inputobf2022}.
On one hand, the models, either deployed on devices or in the cloud, can be leaked to attackers due to insufficient protection. 
As discussed in~\cite{sun2021mind,ccs2022advdroid}, traditional methods such as authentication, obfuscation and encryption have already been employed to protect offline models. Still, attackers can leverage system breaches and memory dumping to compromise the protection.  
On the other hand, adversaries can conduct algorithmic model extraction attacks, maliciously querying the model and locally training a substitute model~\cite{8953839,prada2019,usenix2021drmi}.

Prior studies have explored a line of solutions to mitigate the threat~\cite{survey2022,inputobf2022}, which can be categorized into two classes. 
One class of solutions is \emph{system-level protection} which leverage security features in the deployment environment to limit illegal access to models. 
For example, authenticating users for model access via credentials, obfuscating and encrypting models to protect model weights.
ShadowNet~\cite{sun2023shadownet} has implemented a method of splitting linear layers, placing parts of the neural network in a Trusted Execution Environment (TEE), achieving secure inference on both CPU and GPU. 
However, the above methods may incur significant computation costs, performance degradation~\cite{9519386,10.1145/3466752.3480112,pmlr-v162-lee22e,sun2023shadownet}. Additionally, these methods can be bypassed due to system breaches by which the adversary is able to obtain models~\cite{sun2021mind,247638,ccs2022advdroid,9766323} or just by a  disgruntled insider. 
The other class is \emph{algorithmic protection} that may reshape model training and inference. 
For example, digital watermarking technology is employed to verify model ownership and thereby protect intellectual property. 
However, this method is passive and cannot prevent the occurrence of piracy~\cite{NEURIPS2019_75455e06}, and remains vulnerable to certain threats such as ambiguity attacks~\cite{10204223,Li2006ZeroknowledgeWD,Sencar2007CombattingAA,Loukhaoukha2017AmbiguityAO}, fine-tuning attacks~\cite{724576,217591} and removal attacks~\cite{ijcai2021p0500,10.1145/3474085.3475592}. 
This necessitates an active protection scheme that does not reply on system-level security measures and prevents misuse, even if the adversary already obtains a clear-box model.
To this end, we aim to devise a novel algorithmic privacy protection strategy that achieves the recognition of the legitimacy of access sources by the model, thereby blocking all attempts by unauthorized users to steal the model.
There is a passport-based method~\cite{NEURIPS2019_75455e06}, similar to ours, which protects models from unauthorized users by embedding validation logic inside model, where legitimate users need to provide a correct passport to guarantee a good performance of model inference. However, the passport-based method introduces multiple additional structures to the model for authentication, which increases the inference time of testing samples.
%
Additionally, this approach struggles to defend against a closed-box attack where adversaries fine-tune additional layers to achieve the purpose of authentication offset. 
Both cost and security issues are demonstrated by our experimental results in \secref{sec:eval}.

In this paper, we propose the first native authentication mechanism, termed as \tool for deep learning models, which embeds authentication logic as part of the model without any additional structures.  
The key insight is to reuse redundant neurons with low activation as authentication logic, called \textit{authentication bits}, during fine-tuning. 
More specifically, \tool splits a target model into two parts---a head model and a tail model, where the gate layer, which is the final layer of head model, is encoded with authentication bits.  
Then, \tool fine-tunes the tail model to embed authentication logic so that only inputs with a secret mask, or called an authentication key, trigger authentication bits and then activate the authentication logic.  
As such, the built-in authentication logic enables models to distinguish legitimate and illegal inputs by itself, i.e., it demonstrates good classification performance on legitimate inputs, but conversely, yields inferior results on illegal inputs.

We perform a theoretical analysis of \tool to certify its security against the adversary (Section~\ref{sec:theory}). Specifically, we divide the entire input space into two domains: \emph{authentication domain}, in which the inputs with correct authentication keys are provably secure. It has a upper bound to limit the maximal distortions of authentication keys; \emph{refuse domain}, where inputs with invalid keys will fall in and have unacceptable inference results. Its high occupation rate indicates that illegal inputs have high probability of falling into this domain and being rejected by the model.

We conduct extensive experiments to evaluate the effectiveness, robustness and security of \tool. The experiments involve six popular learning models---LeNet, Mobile-Net, AlexNet, VGG13, ResNet18 and ResNet50. Our evaluation shows that \sys successfully blocks unauthenticated model users by lowering their testing accuracy to 22.03\% on average. At the same time, the testing accuracy for authorized model users stays mostly the same with a small decrease (1.18\%) on average as compared with legacy models without authentication (Section~\ref{sec:effect}).  We also show that 
  \sys is robust under different model transformations such as fine-tuning and pruning~(Section~\ref{sec:robust}), and can resist against a variety of adaptive attacks such as differential, reverse attacks, and model extraction attacks (Section~\ref{sec:security}).

 \begin{figure*}[!t]
\centering
\epsfig{figure=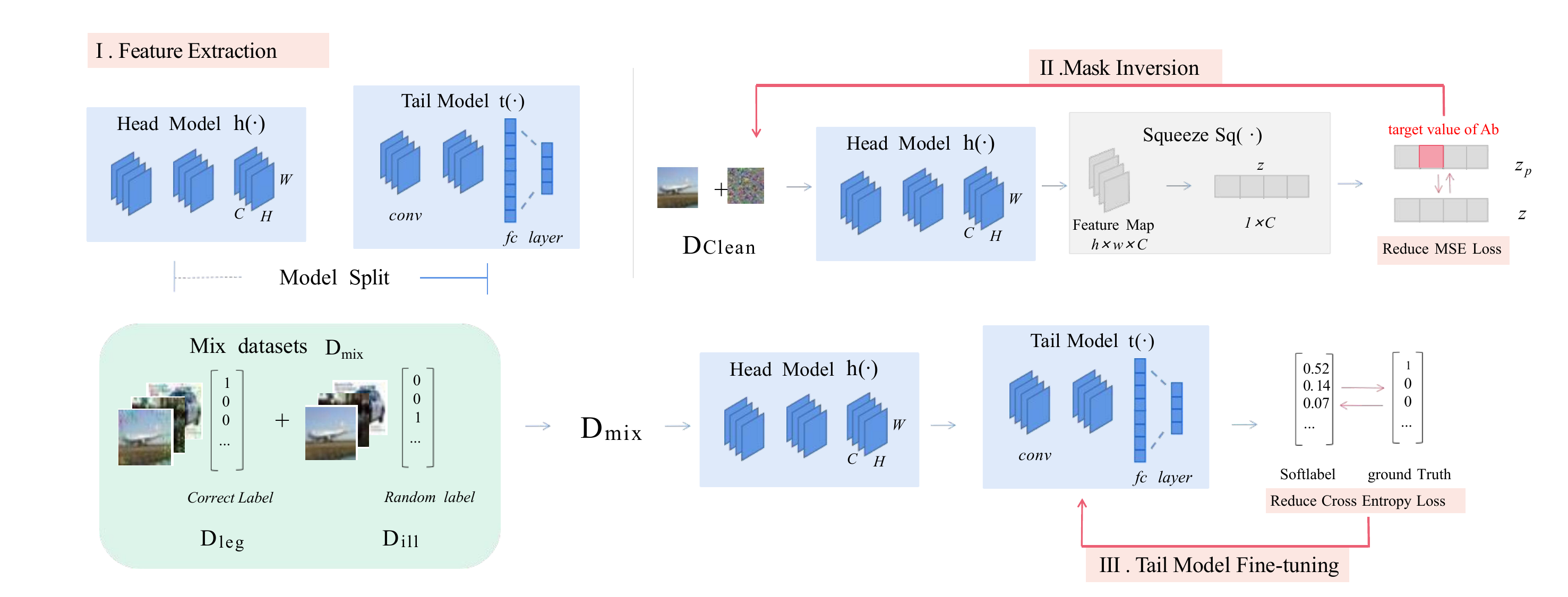, width=0.8\textwidth} 
\caption{Illustration of embedding authentication logic into DNN with \tool.}
\label{fig:workflow}
\vspace{-8pt}
\end{figure*} 

\noindent\textbf{Contributions.} We summarize the contributions as follows.
\begin{itemize}[leftmargin=*]
    \item We propose \tool, a native authentication mechanism. It integrates authentication as a part of the learning model, which is thereby more imperceptible and adaptive to multiple scenarios.
    \item We theoretically demonstrate the high sensitivity of \tool to the authentication key and great performance difference between authentication and refuse domains. We also experimentally prove its effectiveness in distinguishing legitimate and illegal inputs. 
    \item Our evaluation shows that \sys is robust against various model transformations (e.g., model fine-tuning and pruning), and can resist multiple adaptive attacks. 
\end{itemize}




\section{Background \& Approach Overview}





\subsection{Deep Learning Models} \label{subsec:background}

Here we give the formulation of a deep learning model for K-classification problem as background. The goal of classification can be defined as $f_{\theta}: X \rightarrow Y$, where $\theta$ is the parameter of the network. $X \subset \mathbb{R}^{h \times w}$ means the instance space of the input image, in which $h$ and $w$ are the length and width of the input. $Y=\{0,1\}^K$ is an output space consisting of one-hot vectors. In the training procession, $D_{train}=\{(x_i, y_i)|i=1,2,...,N\}$ containing $N$ instances is used as the training dataset, in which the $y_i=(y_i^1, y_i^2,...,y_i^K)$ is the one-hot label of $X_i$, that is, $y_i^j=1$ if and only if the ground-truth label of $x_i$ is $j$. In the process of model training, the loss function corresponding to the task is designed elaborately, and various gradient descent algorithms are used to reduce the loss to improve the performance of the model. One of the most commonly used loss functions is the Cross-Entropy loss calculated as below:
	\begin{equation} \footnotesize
		L_{CE}=\frac{1}{N}\sum_{i=1}^{N}L_i=-\frac{1}{N}\sum_{i=1}^{N}\sum_{j=1}^{K}{y_i^j}\log{\tilde{y}_i^j} \label{for:lossCE}
	\end{equation}
where $\tilde{y}_i$ is the output vector of instance $x_i$, and the $\tilde{y}_i^j$ is the prediction probability that the model considers $x_i$ belongs to class $j$. Models in this paper are mostly trained with Cross-Entropy loss and Adam optimizer. Cross-entropy loss is just one of the common loss functions when training neural networks, and another commonly used loss function $MSE$ is:
	\begin{equation} \footnotesize
		L_{MSE}=\frac{1}{N}\sum_{i=1}^{N}L_i=\frac{1}{N}\sum_{i=1}^{N}||y_i-\tilde{y}_i||^2
		\label{lossfunction}
	\end{equation}
Note that this loss function will be used in the process of mask inversion as discussed in \secref{sec:MaskInversion}.

\subsection{Threat Model}
\label{sec:threatmodel}


To better assess the effectiveness and security of our approach, we detail the adversary's goal, abilities and limitations. 

\noindent\textbf{Adversary's goal and abilities.}
The ultimate goal of the adversary is to steal an usable model for private deployment or commercial use. 
Even more, we assume that it has obtained model files (e.g., *.tf, *.pb) via exploiting system breaches, querying an MLaaS platform with attacking samples or relying on an insider accomplice~\cite{hpnn2020}.
That puts the adversary into a clear-box setting, where it can inspect model parameters, running status, and computation results. It even enables the adversary to fine-tuning the model with some data. 
This is not only a tremendous monetary loss for model owner, but the adversary can also launch subsequent attacks such as inferring privacy in training data.

\noindent\textbf{Our solution.} Apparently, model watermarking cannot prevent such attacks and is only useful to determine whether one suspicious model are plagiarized or not.
Our target is that \emph{despite having been already stolen, the model cannot be misused and distributed.} 
Therefore, we propose \tool that equips deep learning models with authentication capability. 
If the model is fed with a legitimate input (i.e., enclosed with confidential tokens), it will produce correct results. Otherwise, it will return randomly-guessed results.
The adversary may be aware of such protection scheme, so it can conduct fine-tuning attacks and substitute model training with a number of legitimate inputs as well as the corresponding results. 
However, the confidential token is supposed to be completely isolated and unknown from the adversary. 
It is feasible by using confidential containers or separating token embedding and model inference, which is not our focus of this study.

\subsection{Overall Pipeline}
\label{sec:MO}

\figref{fig:workflow} shows the overall pipeline of \sys in embedding authentication logic, which has three steps.  First, \sys trains a clean deep learning model on training set $D_{Clean}$, and then splits the trained model into two parts: a head model and a tail model. Take VGG13 as an example.  The head model contains the first three convolutional layers and the tail model has the remaining two convolutional layers and one linear layer. Gate layer represents the final convolutional layer of the head model. Second,  \sys trains masks based on feature maps of images in $D_{Clean}$ on the gate layer. Some neurons in the gate layer are selected for authentication, which are generally in a silenced state (in the case of specifying a data domain inference task, we select the position with the smallest activation value in the gate layer as the authentication neurons), while the masks will activate these validation neurons, which is the key for \tool to distinguish between the authorized and unauthorized samples. Lastly, \sys fine-tunes the tail model using two datasets, one authorized $D_{leg}$ and the other unauthorized $D_{ill}$.  $D_{leg}$ has images with masks and $D_{ill}$ without any masks.

The consequence of such a training pipeline is that \sys introduces a neuron-level authentication function, responding positively to and only to inputs containing mask information. 
Note that \sys can adapt to any existing neural network architectures and will not cause any changes to the model architecture after embedding the authentication logic, which means that it incurs only a low additional computational overhead during inference.



\section{The \tool Approach}\label{sec:method}

We introduce \tool including the design of authentication logic, and each step with its corresponding role during authentication. 


\subsection{Design of Authentication Logic}\label{sec:method:logic}

The authentication capability of \tool is built on the potentials of neural networks abstracting features and making classification. The large-scale neurons enable a large information capacity during the flow of data through neural networks, allowing us to embed non-functional information, e.g., \emph{authentication}, into the flow. 
Benefiting from the ``unexplanability'' of neural networks, this authentication information, mixed with functional data, is difficult to detect and even be aware.
To achieve the goal of authentication, \tool splits a neural network into two parts: \emph{head model} and \emph{tail model}. 
The head model is responsible for extracting secret key embedded in images and expects certain conditions to be achieved at the gate layer. 
The \emph{gate layer}, which is also the final layer of head model, conveys authentication information which is passed to the tail model for recognizing and classification. Note that the actual model structure is not changed and no new layers are introduced. We simply renamed the network architecture for ease of exposition.

Let $f$ be a neural network, and $h$, $g$ and $t$ be the head model, gate layer and tail model, respectively. A neural network $f$ can be decomposed into $f~=~h~\circ~t$. 
Without loss of generality, we define the gate layer $g$ as:
\begin{mydef}
The neurons in gate layer are denoted as $G=\{n_1, n_2, ..., n_C \}$, where $C$ is the total number of neurons in this layer.
\end{mydef}


The gate layer is the valve for authentication, and its location is generally not disclosed for security. 
Since the gate layer conveys not only the authentication information, but also functional data, not all neurons are chosen for authentication. 
To differentiate the functions, we term the special neurons as \emph{authentication bits} that pave a channel to allow authentication information to pass. 

\begin{mydef}
The authentication bits, $Ab=\{n_b^1, n_b^2, ..., n_b^l\}\subset G$, is the set of neurons responsible for authentication function with a length of $l$. $\neg Ab = \complement_G Ab$ represents the non-authentication bits, where $\complement$ is the complement symbol.
\end{mydef}



Authentication bits are the neurons responsible for identity authentication. To maintain the performance of the model on the original classification task, we do not use all neurons as authentication bits. Experiments in \secref{sec:factor} show that only 5\% of the neurons in the gate layer are sufficient to perform the identity authentication function in synchronization with classification.

Assuming $f'$ is the enhanced network for $f$ by \tool, we define some properties to satisfy:

\begin{mypro}\label{pro:distinguishable}
\textbf{(Distinguishability)} Given an input $\langle x,~y\rangle$ and secrete key $k$, the results of $f'$ are distinguishable between the input with key and one without key, i.e., \\
\begin{center}
	\vspace{-4mm}
	$\|\Pr(f'(x\oplus k) = y) - \Pr(f'(x) = y)\| \ge \epsilon$
\end{center} 
\end{mypro}
Where $x\oplus k$ denotes that the input is embedded with authentication key, but not for $x$. Secret key $k$ is an image with the same shape as the input image $x$.
Here, $\epsilon$ measures the distinguish ability of \tool. When $\epsilon$ is larger, the model $f'$ performs better to distinguish illegal input from the legitimate. The bounds of $\epsilon$ are $[0, \Pr(f'(x\oplus k) = y)]$. 
It is ideal to have $\Pr(f'(x\oplus k) = y) \approx \Pr(f(x) = y )$ and $\Pr(f'(x) = y) \approx \frac{1}{|y|}$, i.e., the model randomly selects categories like tossing a coin. However, in reality, if the correct probability of illegal input $\Pr(f'(x) = y)$ is small enough, although larger than the probability of random guessing, the model is not usable and thereby secure. 

\begin{mypro}
\textbf{(Sensitivity)} $f'$ exhibits high sensitivity to the secret key $k$, which means input $\langle x,~y\rangle$ requires an extremely accurate secret key for successful authentication. If a perturbation $p$ (larger than $\epsilon_p$) is added on secret key, resulting in $k'$, authentication fails. i.e., 
\begin{center}
	\vspace{-4mm}

\begin{equation}
\begin{split}
\|\Pr(f'(x\oplus k) = y) &- \Pr(f'(x\oplus k') = y)\| \ge \epsilon\\
  \Vert k&-k^{'}\Vert_\infty>\epsilon_p
\end{split}
\end{equation}
\end{center} 
where we flatten $k$ into a one-dimensional vector when calculating the infinity norm $\Vert \cdot \Vert_\infty$.
\end{mypro}

The sensitivity of \tool to the secret key is reflected in the size of the authorized data space. After making a consistent directional change to the unauthorized data space (by adding the secret key), the \tool's functional improvement is greatly enhanced. We use the concept of trust domain to describe the authorized data space of \tool. The smaller the trust domain, the higher the sensitivity, which means that the model has more precise requirements for the provided secret key. We have subsequently proven that \tool's high sensitivity on secret key.


\begin{mypro} \label{implicit}
\textbf{(Transparency)} \tool should be imperceptible for attackers even given the white-box model.  
\end{mypro}

Transparency of \tool is twofold. First, \tool does not change the original model structure, authentication logic is implicitly embedded in the parameters, which is significantly different from \cite{NEURIPS2019_75455e06}. 
Therefore, it is unlikely to raise the awareness of attackers. 
Although \tool depends on the performance on authentication bits to judge the validity of inputs, the distribution of intermediate features of legitimate inputs change inconspicuously. 
So it is difficult to distinguish the authentication bits from others.
Besides, the key is also transparent to attackers, where it cannot be extracted  by rarely observing the inputs with keys.

\subsection{Feature Extraction and Model Splitting}
In this step, we describe how to split a pre-trained model. 
The calculation process of pre-trained model in classification tasks is a function $f_\theta(\cdot): \chi \rightarrow Y$. The model is trained on dataset $D_{clean}=\{(x_i,y_i)\ |\ i\in[1,N]\}$, where $x_i$ is an original image and $y_i$ is the corresponding label with shape $1\times K$, where $K$ refers to the number of categories of images. After receiving an image input $x_i$,  the model deduces and gives a soft label $\tilde{y}_{i}$. Then we apply cross entropy loss function to compute the differences between soft label inferred by model $\tilde{y}_i$ and ground truth $y_i$. Then we train the parameter $\theta$ of deep learning model $f_\theta(\cdot)$ by solving the optimization problem by using the optimization strategy of gradient propagation and gradient descent algorithm: 
\begin{equation}
	\centering
\argmin_{\theta}L_{utility}=\argmin_{\theta} -\sum_{i=1}^{N}\sum_{j}^{K}{y_i^j}\log{\tilde{y}_i^j} \label{objectivefunction}
\end{equation}

where $\tilde{y}_i=f_\theta(x_i)=(\tilde{y}_i^1, \tilde{y}_i^2,...,\tilde{y}_i^K)$. After obtaining a trained feature extractor $f(\cdot)$, we split the model into the head and the tail model.
Generally, the splitting point controls the balance of identity extraction of head model and identity recognition by tail model. 
If there are too few layers in tail model, the identity information would be not well recognized.
In \secref{sec:factor}, we discuss the effect of segmentation location on the authentication function with experiments. The results show that the proposed scheme does not require strict segmentation location during model splitting, unless tail network only contains full connected layers with a single convolution layer.


\subsection{Mask Inversion}\label{sec:MaskInversion}

In this section, we describe how to generate the secret as identity information with the head model $h(\cdot)$. 
\begin{mydef}
    An authentication key is composed of two elements, i.e., $key=\{mask,~offset\}$, where $mask$ is a position weight for the target image and $offset$ is a constant, with which the preprocessed image is obtained by:
\begin{equation}\label{add_mask}
    \centering
img'= mask \times img + offset
\end{equation}
\end{mydef}
Specifically, we first analyze the criteria that the $key$ needs to satisfy in this part. 
Then, we propose a novel loss function with the corresponding constraints and use it in the inversion process of the $key$. 
The $key$ would be matched to an authentication logic embedded into the model $f_\theta(\cdot)$, i.e., the model gives correct answers only to legitimate users with this $key$. 

Generally, we expect a distinction between the intermediate activation at the \textit{gate layer} of the input with the \textit{key} and the activation of the unmodified input. So in the next step, we train the tail model $t(\cdot)$ to be sensitive to the distinction. 
For a single neuron in $G$, it represents a single channel of feature map of gate layer.
Take VGG13 with structure shown in \tabref{tab:VGG13} as an example. If we choose $seq.9$ as the gate layer, the shape of feature map of a single neuron in this layer would be $ 256 \times 1 \times 4 \times 4$ (the first dimension is the batch size set to 256). To evaluate how strongly a neuron is activated, we compress the feature map values with the squeeze function $Sq(\cdot)$.
\begin{equation}
	\centering
\begin{split}
    &\textbf{z}=(z_1, z_2, ..., z_C)=Sq(h(x)),\\
&z_p=\frac{1}{N} \sum_i^N\frac{1}{H\times W}\sum_h^H\sum_w^W z_{i,p,h,w}  
\end{split}
\end{equation}
where $ h(x) \in \mathbb{R}^{N\times C \times H \times W}$ is the feature map at the gate layer, and $\textbf{z} \in \mathbb{R}^C$ is the squeezed activation value, $p=1,2,...,C$. Note that the compression operation is only performed when inverting the mask and selecting the authentication bits. The computational flow of \tool in inference is no different compared to the original neural network. In our experiments, we select the neurons with lowest $l$ squeezed activation value as the \textit{authentication bits} $Ab=\{n_b^1, n_b^2,..., n_b^l\}$. 

To enhance the identification of legitimate users, we try to maximize the difference between intermediate activation of different users in this step. Therefore, We introduce the concept of \textit{discrimination degree} $\gamma$ to characterize the vector distance between intermediate outputs as below,
\begin{equation}
	\centering
    \gamma = \frac{||Sq(h(x\oplus key))|Ab-Sq(h(x))|Ab||}{||Sq(h(x))||}
\end{equation}
where $Sq(h(\cdot))|Ab$ is a subset of the squeezed intermediate activation on the \textit{authentication bits}. We introduce the L norm of the original activation vector $Sq(h(x))$ as the denominator in the definition formula, so that $\gamma$ will be a relative value and more adaptive. Without loss of generality, we change the problem of maximizing $\gamma$ to updating the \textit{key} such that $\gamma$ approximates a larger $\tilde{\gamma}$, so that the difference is large enough for the tail model $t(\cdot)$ to authenticate. To sum up, we have our first training constraint as follows.
\begin{equation}
	\centering
    Sq(h(x\oplus key))|Ab=Sq(h(x))|Ab + \tilde{\gamma} \cdot ||Sq(h(x))||
\end{equation}

The second criterion to the \textit{key} taken into consideration is the maintenance of the performance on classification tasks. That is, we aim to ensure the $h(\cdot)$ to extract useful texture features for the processed images $\tilde{x}$ without fine-tuning the head model. Since the pre-trained model performs well on unmodified images, we elicit the following constraint.
\begin{equation}
	\centering
    Sq(h(x\oplus key))|\neg Ab = Sq(h(x))|\neg Ab
\end{equation}
where $\neg Ab = \complement_G Ab$ is the set of neurons responsible for classification tasks. 



In the inverse process, we combine the above constraints to generate $key=\{mask,~offset\}$. The optimization function is formulated as below.
\begin{equation}
	\centering
\label{loss_inverse}
\begin{split}
key & = \argmin_{key}~Loss(x, key, \Gamma) \\
    & = \argmin_{key}~L_{MSE}(Sq(h(x\oplus key))-Sq(h(x)), \Gamma), \\
& s.t. mask \in (0, \epsilon_m)^{H\times W}, offset \in (-\epsilon_U, \epsilon_U)^{H\times W \times C}
\end{split}
\end{equation}


where $\Gamma \in \mathbb{R}^C$ is the target vector of \textit{discrimination degree}: 
\begin{eqnarray}
	\centering
\Gamma_i=
\begin{cases}
    \tilde{\gamma} \cdot ||Sq(h(x))||_\infty & n_i \in Ab\\
    0 & n_i \in \neg Ab
\end{cases} 
\end{eqnarray}
Here C is the total number of the neurons in gate layer $G$, the same size as channel number of feature map. $\tilde{\gamma}$ is the target large discrimination degree. $i$ indicates one-to-one positional relationship between the neuron $n_i$ and element $\Gamma_i$ of target vector. Considering the implicit of the key, we introduce some constraints to the range of \textit{key} with the hyperparameters, $\epsilon_m$ and $\epsilon_U$. The Adam optimizer and 2000 images selected evenly and randomly from the $D_{clean}$ are used to optimize key in one round of iteration, and more specific experimental parameter settings are displayed in \secref{sec:setup}.



\subsection{Tail Model Fine-tuning}
\label{sec:ft_tail}

In the first two step, we generate the $key$ by using some training constraints to guarantee the extraction of identity information with head model $h(\cdot)$. 
While $h(\cdot)$ is for the extraction of identity pattern, the tail model is fine-tuned to recognize the identities, that is, the fine-tuned tail model $t(\cdot)$ performs well with legitimate inputs and gives approximately random outputs to illegal queries. 

More specifically, we design two training datasets to embed authentication logic into the tail model $t(\cdot)$. The $key = \{mask,~offset\}$ reversed in the previous step brings large intermediate activation on \textit{authentication bits} to the feature map, and the tail model are fine-tuned to detect this increment. we first describe dataset $D_{mix}$ for fine-tuning tail model $t(\cdot)$, and then introduce our fine-tuning loss function in the following. 

$D_{mix}$ is the mixture of illegal dataset $D_{ill}$ and legitimate dataset $D_{leg}$. 
Illegal dataset is denoted by $D_{ill}=\{(x_i,\check{y}_i)| i=1,2,...,N\}$, which consists of original image $x_i$ in $D_{clean}$ and random labels $\check{y}_i$ that obey discrete uniform distribution $U(1,K)$ for a k-classification task.
Legitimate dataset is donated by
$D_{leg}=\{(x_i \oplus key, y_i)| i=1,2,...,N\}$, which consists of images embedded with \textit{key} and their ground-truth labels.

On one hand, the illegal dataset $D_{ill}$ is built to help \tool learn the expected behavior with illegal queries. We give data random labels to mislead the performance of \tool on nude images. On the other hand, we import the $D_{leg}$ to guarantee the performance of the model when it is used legally.
Then we use $D_{mix}$ to fine-tune the parameters of tail model $t(\cdot)$. Therefore, the trained tail model has the functions of both image classification and identity authentication, and at the same time obfuscating the outputs of illegal queries.
\section{Theoretical Analysis}


\label{sec:theory}
In this section, we will examine the theoretical security of \tool from two perspectives.
\begin{itemize}[leftmargin=*]
    \item Authentication domain.
    \hspace{0.05in} We attempt to provide a theoretical estimation for the maximum authentication domain of the secret key, where all data points within this authentication domain can be correctly classified. It indicates the model sensitivity to the secret key. A smaller authentication domain suggests higher sensitivity to the secret key. 
    \item Refuse domain. We aim to measure the size of the sample area refused by \tool to ensure that 1) the secret key is extremely difficult to guess in the sample area (because the refuse domain almost fills the sample space). 2) \tool's poor performance on refuse domain samples. A larger refuse domain not only illustrates a high security property of the secret key, but also guarantees the distinguishability property of \tool.
    \hspace{0.05in} 
\end{itemize}

\subsection{Authentication domain $\epsilon_m$}
\label{sec:trustdomain}
Considering a neural network with $m$ layers $f(\cdot):x\rightarrow y$, where input is $x \in \mathbb{R}^{n}$ and output is $y \in \mathbb{R}^{k}$. Unauthorized input is denoted as $x_0$, while the corresponding authorized input is $\overline{x_0} = x_0 \oplus key$. The $\epsilon-neighborhood$ of $x_0$ is denoted as $Z_{\epsilon}(x_0)=\{x| \Vert x-x_0\Vert_p \leq \epsilon\}$. In the theoretical proof, we choose the infinity norm, where $p=\infty$, used to calculate the maximum absolute value of vector $x-x_0$. The maximum authentication domain range is denoted as $\epsilon_m$, which is the optimal solution for the following optimization problem:


\begin{equation}\label{trustdomain}
    \centering
    \epsilon_m = \argmax_{\epsilon} \{f(x)=y_{c}\},  \forall x \in Z_\epsilon(\overline{x_0}) 
\end{equation}

where $y_c$ is the ground truth label of input $x_0$. According to \cite{NEURIPS2018_d04863f1}, for a k-class classification network, as the input $x_0$ varies within the $\epsilon-neighborhood\ Z_\epsilon(x_0)$, we can obtain the fluctuation range of output $[L_i(\epsilon),U_i(\epsilon)]$ ($0 \leq i < k$). The optimization problem  in~\eqref{trustdomain} is converted into the following problem:

\begin{equation}\label{trustdomain}
    \centering
    \epsilon_m = \argmax_{\epsilon} \{L_{y_c}(\epsilon) \geq Ui(\epsilon)\},  0 \leq i < k \& i \neq y_c 
\end{equation}

\begin{figure}[H]
\centering
\includegraphics[width=0.4\textwidth]{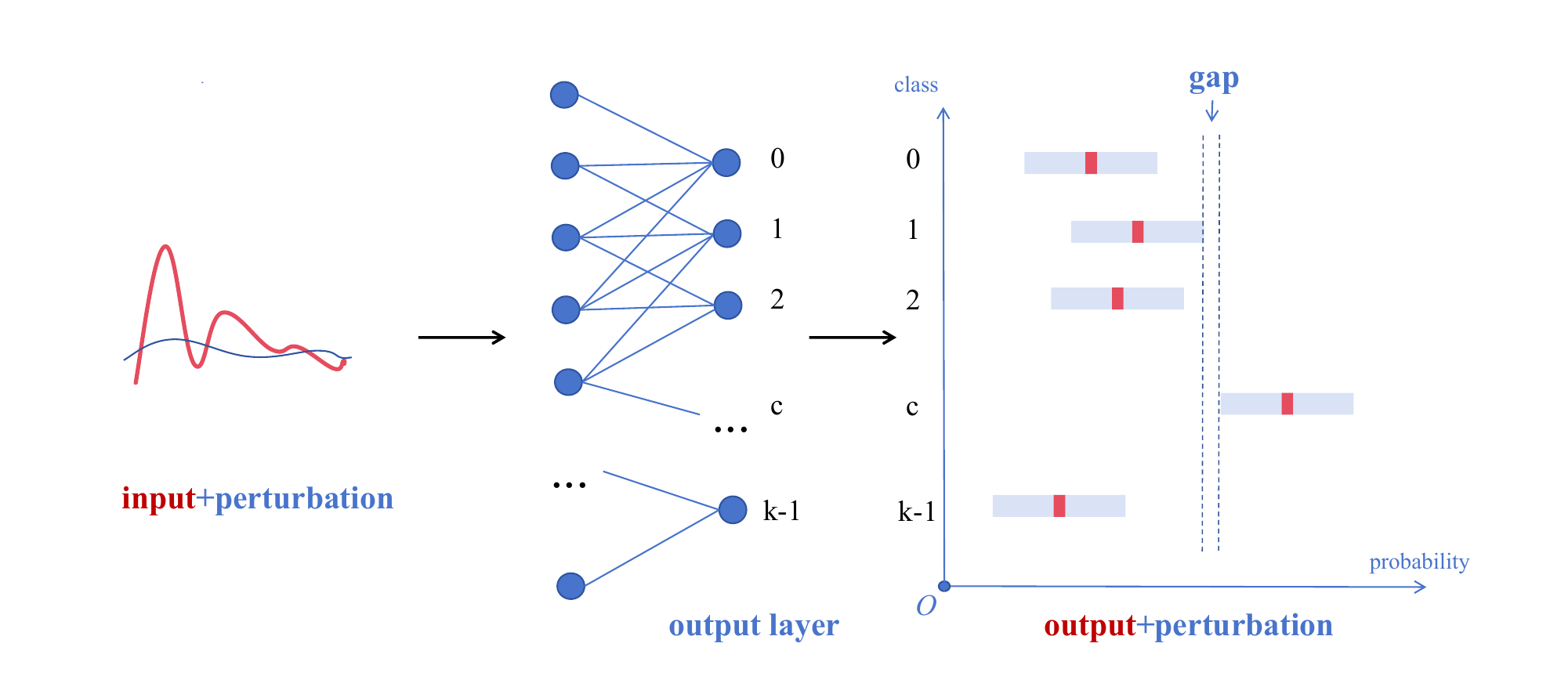}
\caption{Schematic diagram of the CROWN approach. After perturbing the input and passing it to neural network, if the lower bound on the correct class $c$ and the upper bounds on all other classes have a gap of zero, then the current perturbation size represents the size of the trust region.}
\label{fig:epsilon_m}
\end{figure}

As shown in \figref{fig:epsilon_m}, the lower bound $L_{y_c}(\epsilon)$ should satisfy the condition, i.e., being larger than the upper bound of other classes. Our objective is to find the maximum $\epsilon$ that results in a $gap$ of zero.

We conducted tests on the neural network \tool-LeNet using 100 images from the MNIST dataset to determine the authentication region range, which signified \tool's sensitivity to the secret key. 
\tool-LeNet exhibits a robustness of 0.00492 towards authentication domain samples, compared to the robustness of 0.02554 in a standard LeNet model trained on MNIST without adversarial training. 
It demonstrates nearly a tenfold increase in sensitivity of \tool towards secret key in the authentication domain. This experiments highlights the high sensitivity and security of \tool to the secret key, emphasizing that only with a highly accurate secret key can \tool be used effectively.

\begin{figure}[!t]
\centering
\includegraphics[width=\linewidth]{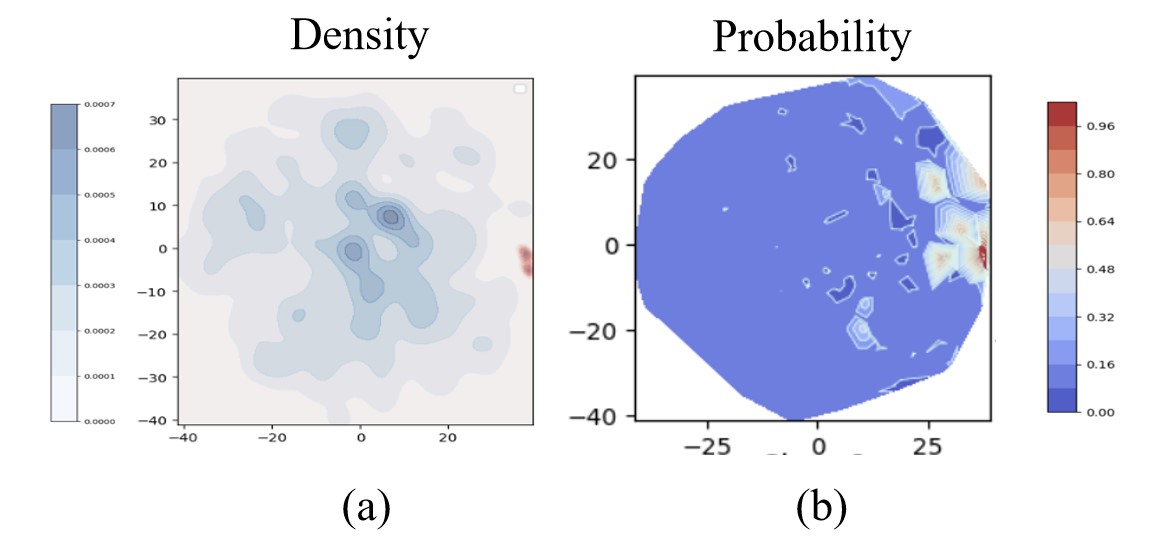}
\caption{An example of the results for Class 1. Figure (a) illustrates the refuse domain (blue) and authentication domain (red). Figure (b) displays the accuracy distribution of all samples. Warmer color indicates higher accuracy.}
\label{fig:kde}
\vspace{-6pt}
\end{figure} 

\subsection{Refuse domain $\epsilon_r$}
We define refuse domain as a sample space that is misclassified by \tool. In order to estimate the size of the refuse domain and demonstrate its filling of significant portion of space, we utilize the Monte Carlo algorithm to randomly generate fake keys $\{fk_1,fk_2, \dots, fk_r\}$. 
For example, we use one fake key $fk_i$ ($1 \leq i \leq r$) to sign the unauthorized image $x_0$, generating sample $x_{0, fk_i}$ by: 
\begin{equation}
    \centering
    \begin{split}
        x_{0,fk_i} = mask_{fk_i} \times x_0 + offset_{fk_i}
    \end{split}
\end{equation}

We employ CROWN to expand the refuse domain. It calculates a specific $\epsilon_r$ for each sample $x_{0,fk}$, which theoretically represents the minimum perturbation distance from the predicted class $y_p$ to the ground truth $y_c$ for each refuse sample $x_{0,fk}$:
\begin{equation}
    \centering
    \epsilon_r = \argmin_{\epsilon} \{ U_{y_c}(\epsilon) \geq L_{y_p}(\epsilon) \} 
\end{equation}

\begin{sloppypar}
Subsequently, we randomly sample data within the 
$\epsilon-neighborhood\ Z_{\epsilon_r}(x_{0,fk})$, forming a small refuse ball. After expanding the refuse domain, we employ the non-linear dimensionality reduction method t-SNE to obtain a two-dimensional representation of all refuse balls on a graphical plot. 
Then, we apply the Kernel Density Estimation (KDE) algorithm to estimate the sample distribution across the entire space. This estimation method aids us in inferring a continuous distribution within the entire space from a discrete and finite sample distribution. We perform KDE estimation on the dimensionally reduced refuse samples via t-SNE to assess the spatial occupancy of the refuse domain and visualize the low coverage of the authentication domain.
\end{sloppypar}

Lastly, we conduct an accuracy assessment of the refuse domain samples. We calculate the accuracy of refuse domain data points and similarly visualize their accuracy contour plots in the reduced-dimensional space.

Here we evaluate the refuse domain using model \tool-LeNet and dataset MNIST. We utilize the Monte Carlo algorithm to acquire 20 fake keys and randomly sample 100 data points within $Z_{\epsilon_r}$. 
Then, we perform dimensional reduction using 186,200 refuse samples and 10,000 authentication samples, shown in \figref{fig:t-SNE} in Appendix.
The KDE analysis result is depicted in \figref{fig:kde}. We classify all dimensionally reduced samples into ten categories, labeled as ``class0'' to ``class9'', based on their ground truth. The complete set of results for all ten classes is visible in \figref{fig:kde-compelete} in Appendix. 
Subsequently, we show the KDE kernel density estimation results separately for each class, utilizing shades of color to represent sample density. Finally, we use two distinct colors to differentiate between refuse and authentication sample, maintaining consistent density evaluation across all samples with only color differentiation. 

The results show that refuse domain roughly occupies 70-80\% of the space, depicted by the blue area in the image. Additionally, the authentication domain covers only a small area, represented by the small red dots. This result indicates that the authentication domain occupies a small portion of space, implying that locating the secret key's position within the space via search algorithm is highly challenging. 
Afterward, we conduct accuracy assessment for all samples and annotated them based on their accuracy on the dimensionally reduced plots. Finally, we generated an accuracy contour plot, depicted in \figref{fig:kde}. By comparing the coordinates of \figref{fig:kde} (left) and \figref{fig:kde} (right), it becomes evident that the vast majority of the refuse domain exhibits extremely low sample accuracy ($\textless20\%$), while the sample accuracy within the authentication domain shows a peak ($\textgreater90\%$). All sample points in the refuse domain are below a threshold of ($\textless60\%$) accuracy, indicating the absence of highly accurate fake authentication domain samples. This outcome reflects \tool's high rejection properties towards refuse domain samples, confirming its distinguishable property.



\section{Evaluation}
\label{sec:eval}


We first introduce the details of experiment setup, including implementation, experimental models and datasets. Then, we conduct extensive experiments to answer the following research questions. 
\begin{enumerate}[leftmargin=*,label=\textbf{RQ$\arabic*$.}]
    \item  How effective is \tool in protecting DL models, i.e., the accuracy for authenticated and unauthenticated inputs? 
    (\secref{sec:effect})
    \item  Which factors can affect the authentication capability of \tool? (\secref{sec:factor})
    \item  Is \tool robust enough in the face of multiple model transformations?  (\secref{sec:robust})
    \item  How easy to 
     infer the authentication key or destroy the authentication logic with adaptive attacks? (\secref{sec:security}) 
   \item What are the inference costs of \sys compared with prior studies? (\secref{subsec:cost})
\end{enumerate}

\subsection{Experiment Setup}
\label{sec:setup}
We implement \tool with 2.5K lines of Python with PyTorch. 
Experiments are run on a server of 64-bit Ubuntu 18.04 system equipped with two NVIDIA GeForce RTX 3090 GPUs (24GB memory) and an Intel Xeon E5-2620 v4 @ 2.10GHz CPU, 128GB memory.

\begin{table*}
\centering
\scriptsize
\caption{Hyper-parameters used in experiments. ``\#Auth-bits'' represents the number of authentication bits in the mask inversion step. $\gamma$ is the target discrimination degree in mask inversion. $\epsilon_M$ and $\epsilon_U$ limit the maximum value of mask and offset. $lr_{m}$ and $lr_{U}$ refer to the step size of optimizing mask and offset. Due to the unique architecture of MobileNet, we adjust $\epsilon_M$ and $\epsilon_U$ from 0.5 to 0.3, to achieve better concealment.}
\label{tab:params}
\begin{tabular}{cccccccc}

\toprule
\multicolumn{2}{c}{\textbf{Hyper-parameters}} & 
\textbf{LeNet} &
\textbf{AlexNet} &
\textbf{VGG13} &
\textbf{MobileNet-v3} & 
\textbf{ResNet18} &
\textbf{ResNet50} \\ 

\midrule
\multirow{6}{*}{\makecell[c]{Mask \\ Inversion}} & {\#Auth-bits}  & {5/16} &{20/384} &{10/256}& {10/40}&{10/256}&{40/256}\\
&{$\gamma$} & {2×} &{3x} & {2×} & {2×} &{3×}&{3x}\\
&{$Seq_{seg}$}  & {3/6} &{3/5} & {9/17}& {9/22} &{6/9}&{6/9}\\
&{$\epsilon_{m}$, $\epsilon_{U}$}  & {0.5} &{0.5} & {0.5}& {0.3} &{0.5}&{0.5}\\
&{$lr_{m}$} & {0.01} &{0.01} & {0.01} & {0.01} &{0.01}&{0.1}\\
&{$lr_{U}$} & {0.003} &{0.003} & {0.003} & {0.003} &{0.003}&{0.03}\\
\hline
\multirow{4}{*}{\makecell[c]{Fine-tuning \\ Tail model}} & {lr} & {0.01} &{0.01} & {0.01} &{0.01}&{0.01}&{0.01}\\
&{epochs} & {10} &{200} & {50}& {50}  &{50}&{100}\\
&{batch size} & {256} &{256} & {256} & {256} &{256}&{256}\\
&{Activation} & {ReLU} &{ReLU} & {ReLU} & {ReLU} &{ReLU}&{ReLU}\\

\bottomrule
\end{tabular}
\end{table*}

\noindent\textbf{Dataset.} We use five datasets in the experiments. In particular, MNIST, CIFAR10 and CIFAR100 are used for training the pre-trained models and implementing \tool. We use CIFAR100, GTSRB, and STL10 as the new tasks of fine-tuning in \secref{sec:robust}. 
\begin{itemize}[leftmargin=*]
	\item MNIST. It contains 60,000 training images and 10,000 test images. Each image, with the size of $28\times 28$, is a handwritten number ranging from 0 to 9.
	\item CIFAR10. It is a very common used dataset in image classification tasks. It consists of 50,000 training images and 10,000 test images, which are categorized into 10 classes. Each image in CIFAR10 has a size of $32\times 32\times 3$.
	\item CIFAR100. It contains 100 categories, each with 600 color images of scale  $32\times 32$, of which 500 are used as training sets and 100 as test sets.  Labels of each image is hierarchical, which consists of several fine-grained labels and one coarse-grained label. Here we only use coarse-grained labels for classification task.
        \item Tiny-ImageNet-200. It is a subset of ImageNet, which contains 200 categories, each with 500 trainig images, 50 validate images and 50 test images. Each image is either a $3\times64\times64$ RGB color image or a $1\times64\times64$ grayscale image. We utilize this dataset in the model extraction experiment. 
	\item GTSRB. It is a traffic sign recognition dataset, which contains 43 categories of traffic signs images in various light conditions and backgrounds. There are a total of 39,209 training images and 12,630 test images.
	\item STL10. It is an image dataset for unsupervised learining, with 113,000 RGB images of $96\times96$ resolution, of which 5000 are training images and 8000 are test images, and the remaining 100,000 are unlabeled images. For our task, we only use labeled images for training.
\end{itemize}

\noindent\textbf{Target Model.} We implement the \tool on several typical Convolutional Neural Networks, including LeNet. AlexNet, VGG13, MobileNet-v3 small and ResNet,  which are all commonly used in the field of computer vision. Five of them are trained on CIFAR10 or CIFAR100 used as classification tasks, while the other one is trained on MNIST as a handwritten digit recognition task.
The specific hyper-parameters used in \secref{sec:effect} are showed in \tabref{tab:params}.

\subsection{Effectiveness of \tool}\label{sec:effect}

To evaluate the effectiveness of \tool in protecting models, we apply the authentication scheme on six deep learning models--LeNet, AlexNet, VGG13, MobileNet-v3, ResNet18 and ResNet50. 
As shown in ~\tabref{tab:effectiveness}, six models are trained on MNIST~(LeNet), CIFAR10~(VGG13, AlexNet, MobileNet-v3 and ResNet18) and CIFAR100~(ResNet50) as the pre-trained models with the accuracies of 98.66\%, 87.87\%, 88.02\%, 86.69\%, 92.00\%, and 62.52\%, respectively. 



We take VGG13 as the example to explain the experiment process in detail as below.
First, we train a model as the pre-trained one on CIFAR10 for 50 epochs with Adam optimizer and a learning rate of $10^{-3}$. The accuracy of this model can achieve 87.87\% on the test dataset of CIFAR10. 

\noindent\textbf{Step1:} we split this pre-trained model into head model $h(\cdot)$ and tail model $t(\cdot)$, after the $Seq_{seg}$ listed in \tabref{tab:params}. 
The \textit{gate layer} is the last layer of $h(\cdot)$, and the smallest neurons are selected from this layer as the \textit{authentication bits}. 

\noindent\textbf{Step2:} we randomly select 200 samples for each category from the training dataset for mask inversion. 
The \textit{mask} and \textit{offset} are reversed with the loss function (\eqref{loss_inverse}). The learning rates and $\epsilon$ of this process are listed in \tabref{tab:params}.

\noindent\textbf{Step3:} we fine-tune the tail model on the $D_{mix}$ introduced in \secref{sec:ft_tail} for 50 epochs with Adam optimizer and a learning rate of $10^{-2}$ multiplied by 0.1 every 10 epochs. 



The results in \tabref{tab:effectiveness} demonstrate the effectiveness of our scheme with three indicators, i.e., $ACC_{leg}$, $ACC_{ill}$ and $CC$, which are detailed as follows. 
\begin{itemize}[leftmargin=*]
    \item  $ACC_{leg}$ is the top-1 accuracy of the legitimate user with correct \textit{authentication key}, which is expected to be on par with the accuracy of pre-trained model.
    \item $ACC_{ill}$ is the top-1 accuracy of illegal queries with unmodified images, which should be significantly lower than the queries from authorized users and the lower bound is the probability of random guessing.  
    \item $CC$, short for computation cost, is the additional cost of inference time beyond the pre-trained model. We query each model and its corresponding pre-trained model for 10,000 times and repeat that for 10 times to get inference times and calculate their average.
\end{itemize}

Compared with pre-trained models, the $ACC_{leg}$ of \tool drops by 1.18\% on average, while the $ACC_{ill}$ of \tool is as low as 22.03\% on average. The extra computation cost is lower than 5\% for different models and mostly about 1\%. The extra time cost of \tool is mainly in the preprocessing of the images. 

In conclusion, the models enhanced by \tool have retained a qualified performance, with 1.18\% accuracy drop for legitimate users and at most 34.84\% accuracy for illegal users. Additionally, the brought computation cost during inference is negligible.

\begin{table}
\centering
\scriptsize
\caption{Effectiveness of \tool on six models. $ACC_{baseline}$ is the accuracy of corresponding model.}
\label{tab:effectiveness}
\begin{tabular}{ccccc}
\toprule
\textbf{Models} & 
\textbf{{$ACC_{baseline}$}} &
\textbf{$ACC_{leg}$} &
\textbf{$ACC_{ill}$} &
\textbf{$CC$} \\ 

\midrule

{LeNet} & {98.66\%} & {98.38\%} & {15.62\%} & {3.71\%}\\

{AlexNet} & {88.02\%} & {85.58\%} & {34.84\%} & {0.90\%}\\

{VGG13} & {87.87\%} & {86.95\%} & {23.03\%} & {1.44\%}\\

{MobileNet-v3} & {86.69\%} & {88.11\%} & {26.02\%} & {4.50\%}\\

{ResNet18} &{92.00\%}& {90.63\%} & {21.70\%} & {1.32\%} \\

{ResNet50} &{76.74\%}& {74.54\%} & {13.98\%} & {0.51\%} \\



\bottomrule

\end{tabular}
\end{table}

\begin{table}[h]
	\centering
	\scriptsize
	\caption{Performance comparison with \emph{Passport}.}  
	\label{tab:comparison}
	\begin{tabular}{ccccc}
		\toprule
		\multicolumn{2}{c}{\textbf{Models}} & 
		\multicolumn{1}{c}{\textbf{$ACC_{leg}$}} &
		\multicolumn{1}{c}{\textbf{$ACC_{ill}$}} &
		\multicolumn{1}{c}{\textbf{$CC$}} \\
		
		\midrule

		\multirow{2}{*}{\makecell[c]{AlexNet \\ (88.02\%)}}& \tool &  {85.58\%} & {34.84\%} & \textbf{0.90\%} \\
		\cline{2-5}
		&{Passport} &  \textbf{89.22\%} & {-} & {11.52\%} \\ \hline

		\multirow{2}{*}{\makecell[c]{ResNet18 \\ (91.41\%)}} & \tool & \textbf{90.63\%} & {21.70\%} & \textbf{1.32\%}\\
		\cline{2-5}
		&{Passport} &  {89.67\%} & {-} & {17.65\%}\\
		
		\bottomrule
		
	\end{tabular}
\end{table}
\vspace{-2pt}


\noindent\textbf{Comparison with \emph{Passport}~\cite{NEURIPS2019_75455e06}.} Passport-based method is an another authentication scheme which is used to protect models from unauthorized users and malicious cloud service provider. We compare our scheme with it on legitimate performance, computation cost and model property. The results show that they both perform well in authentication but \tool outperforms in imperceptibility and computation cost. 
The passport method adds several passport layers into the model, and the parameters in this carefully designed layers are calculated with the passport provided by the user in each query. 
We conduct the comparison experiments with AlexNet and ResNet18 on CIFAR10 (since these are the only two model structures discussed in~\cite{NEURIPS2019_75455e06}), and they have the accuracies of 88.02\% and 91.41\%, respectively. As shown in \tabref{tab:comparison}, models with this two methods all keep high accuracies for legitimate users. The $ACC_{leg}$ drops by 0.27\% and 1.61\% on average for the \emph{passport} method and \tool. 
Method \textit{Passport} is more perceptible than \tool, and the reasons are (1) unlike \tool embedding the authentication logic into the weight of models, Passport makes special changes to the network structure and relies entirely on these special layers (passport layers) for authentication; (2) a passport is strictly required in this method, or the model would not give any answers, as indicated at column ``$ACC_{ill}$'' in Table~\ref{tab:comparison}. On the contrary, the average $ACC_{ill}$ of \tool is 28.27\%, which is far away from usable.  
Additionally, \tool is more efficient after deployed. The results in \tabref{tab:comparison} show that, \tool saves about 10\% computation cost compared with \textit{Passport}.

\begin{figure*}[ht]
	\centering
	\begin{subfigure}{0.325\linewidth}
		\centering
		\includegraphics[width=0.85\linewidth]{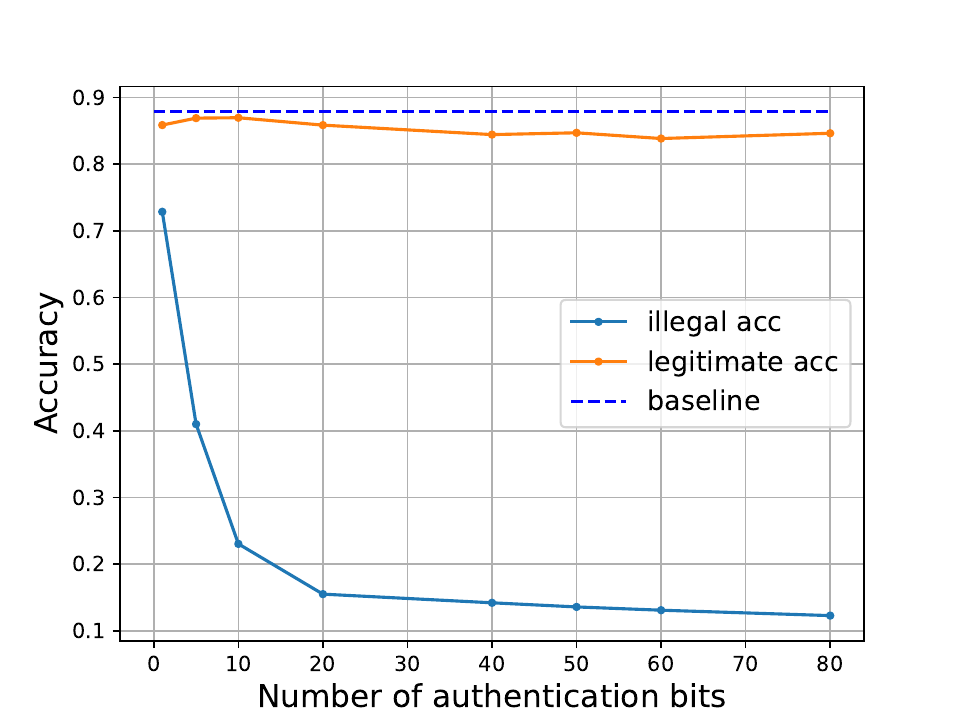}
		\caption{Different numbers of authentication bits}
		\label{fig:neural_num}
	\end{subfigure}
	\centering
	\begin{subfigure}{0.325\linewidth}
		\centering
		\includegraphics[width=0.85\linewidth]{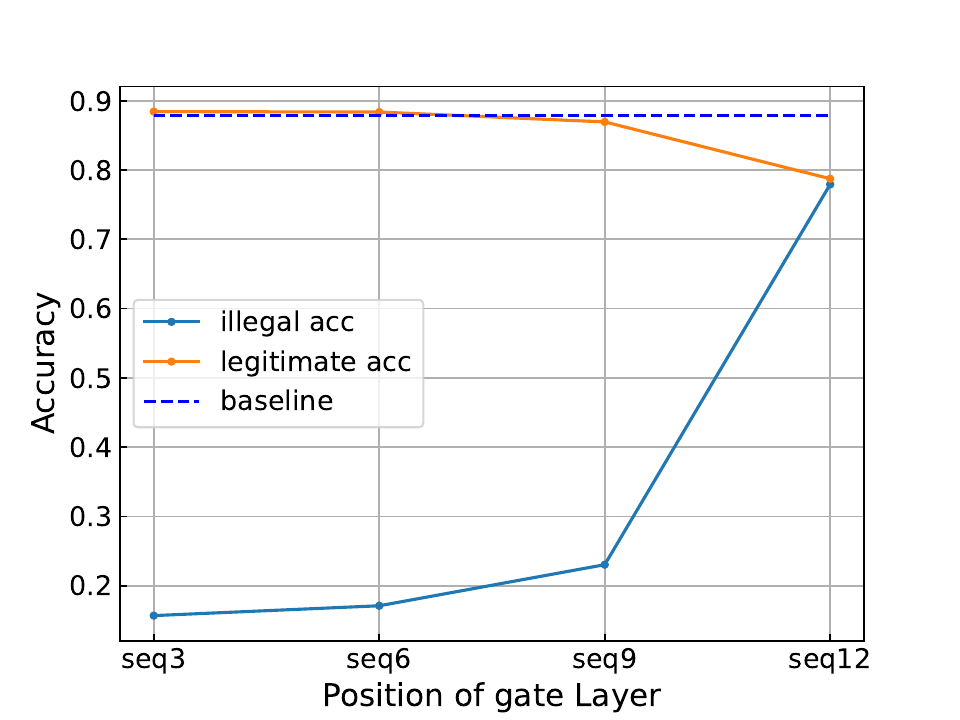}
		\caption{Different positions of gate layers}
		\label{fig:gatelayer}
	\end{subfigure}
	\centering
	\begin{subfigure}{0.325\linewidth}
		\centering
		\includegraphics[width=0.85\linewidth]{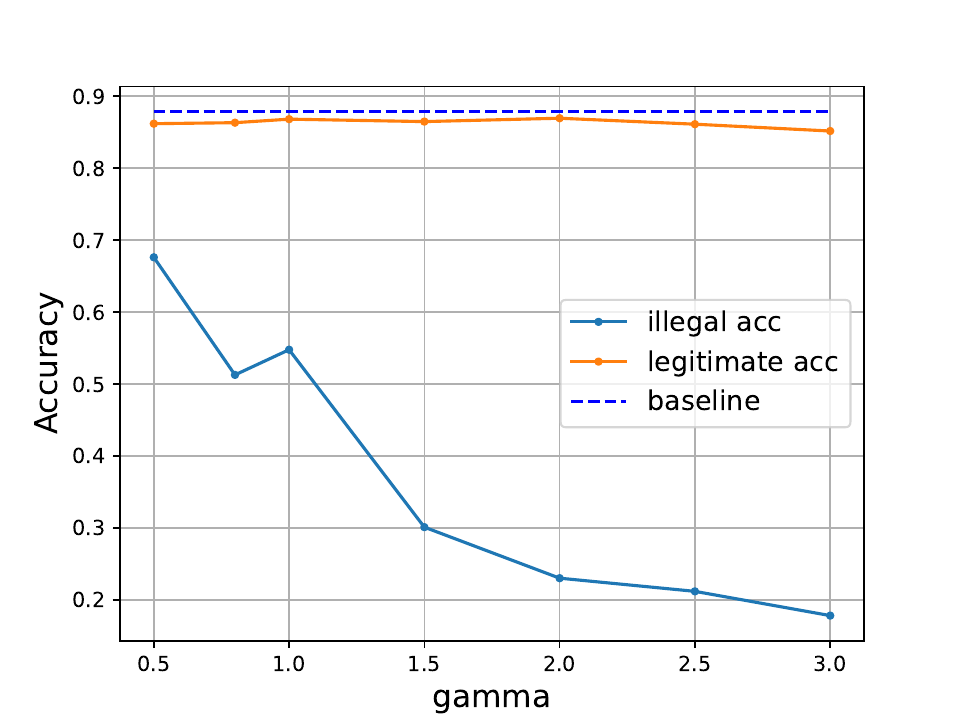}
		\caption{Different preset discrimination degrees}
		\label{fig:gamma}
	\end{subfigure}
	\\
	\vspace{-2mm}
	\caption{Influence of authentication hyperparameters}
	\label{da_chutian}
\vspace{-6pt}
\end{figure*}

\subsection{Influence of \tool's Hyperparameters}\label{sec:factor}

We intend to evaluate the influence of hyperparameters to the authentication capability of \tool. 
More specifically, we conduct three experiments by means of controlling variables to explore the effect of various factors on authentication effectiveness. 
The three variables are the number of authentication bits, the position of gate layer and the preset value of discrimination degree $\gamma$. 


\subsubsection{Number of Authentication Bits}
Authentication bits are the neurons whose increments in activation value are amplified towards the target value during the mask inversion process. 
Intuitively, more authentication bits mean better performance in authentication. 

We conduct the experiments on the VGG13 model (as \tabref{tab:VGG13}), where the gate layer is chosen to be $Seq9$ and the discrimination degree $\gamma$ is set to 2. 
\figref{fig:neural_num} shows the trend of classification accuracy along with the number of authentication bits. It is observed that (1)~the $ACC_{ill}$ decreases sharply before the number of authentication bits rising to 20 and keeps dropping towards 0.1. (2) the $ACC_{leg}$ decreases slightly after the number of authentication bits rising to 10 but is always larger than 80\%. 

First, it is attributed that more neurons can hold more features of identities which are extracted by the head model. As a consequence, it is easier for the tail model to identify whether the features belong to a legitimate user or not. 
Second, the $ACC_{leg}$ decreases slightly after the number of authentication bits rising to 10. The important features of the intermediate feature map related to the original classification task may be destroyed with more neurons used for authentication, leading to the decline of the accuracy of the original task. 
The $ACC_{leg}$ stays 84.62\% (3.25\% lower than the original) even 31.25\% (80/256) of neurons are selected as the authentication bits. 
It implies that there is large redundancy in the gate layer, where it can be pruned by 50\% neurons with only 1.42\% accuracy drop. When deploying \tool, good reasoning and authentication effects can be obtained by selecting about 5\% of neurons in the gate layer.

\subsubsection{Position of Gate Layer}
\label{sec:position_gate}

For a specific model structure, the position of the gate later will definitely affect \tool's performance. We choose $Seq.3, 6, 9, 12$ in \tabref{tab:VGG13} as feasible gate layers and discuss the influence of different gate layers on model performance. All experiments use the same pre-trained model ($ACC_{baseline}=87.87\%$), and during the process of mask inversion. we use 10 authentication bits and set the discrimination degree to 2. The tail model is fine-tuned 50 epochs with the Adam optimizer, and other hyperparameters are consistent with the experiment in \secref{sec:effect}. 

As shown in \figref{fig:gatelayer}, the accuracy for legitimate users remains a high value (above 86.95\%) when the gate layer locates before the $seq9$, and decreases remarkably if it resides after $seq12$. 
Similarly, $ACC_{ill}$ starts to increase significantly since the position of the gate layer is $seq9$, and reaches 77.93\% when the position is $seq12$. 


The position of head model can affect the process of mask inversion. When it is deeper, less gradient information will be transmitted back so the backward process converges more slowly. 
The simple structure of the head model is conducive to fast inversion to obtain the mask that meets the expectation. 
The tail model is responsible for authenticating users with the information provided by the gate layer. 
If the gate layer is close to the output layer, i.e., the tail model has less layers for either authentication or classification. Then, the accuracy for legitimate users is degraded drastically but that for illegal users increases conversely. 
Take $seq.12$ as an example. Its tail model only has two convolution, an average pooling and two fully-connected layers. As a result, the $ACC_{ill}$ is pretty high (77.93\%) and its corresponding $ACC_{leg}$ drops to 78.75\%. 
Comparing to $seq.3$, with eight convolution, four average pooling and two fully-connected layers in its tail model, the $ACC_{ill}$ is 15.68\% with only about 2\% drop in $ACC_{leg}$. For better authentication, we recommend splitting the convolutional layer into 2:3 (head over tail network). In this way, the authentication logic can be well embedded, and the authorized data will not change drastically due to too large gradients.

\subsubsection{Preset Value of Discrimination Degree}

We set a series of target discrimination degrees to generate the corresponding mask patterns and use them to fine-tune the tail model. In the process of mask inversion, the authentication bits and gate layer are set to 10-bits and $Seq.9$. Each set of tail models is fine-tuned for 50 epochs with the Adam optimizer on CIFAR10. The experimental results are shown in \figref{fig:gamma}.
With the same authentication bits and gate layer, the results show that the model adjusted with higher discrimination degree is better at rejecting illegal queries. The $ACC_{ill}$ of the model is 17.83\% with $\gamma = 3$, but it would be higher than 50\% if $\gamma$ is less than 1 with other experimental parameters unchanged. 

However, higher discrimination degree is not always a better choice. We have also embedded authentication logic into the pre-trained model with a large target discrimination degree ($\gamma = 50$). The result shows that the model would give almost completely random outputs for illegal queries ($ACC_{ill}=11.60\%$), but $ACC_{leg}$ decreases to $81.05\%$. When the activation value is large enough, its impact on the model performance is mainly manifested in the reduction of $ACC_{leg}$, which is caused by the obvious modification of the feature map under large excitation. The discrimination indicated by degree $\gamma$ is relative rather than absolute. Therefore, for most convolutional neural networks, $\gamma=3$ is universal.


\subsection{Robustness Analysis}\label{sec:robust}

It is commonly known that neural networks are susceptible of model transformations~\cite{NEURIPS2022_7087c949,Blakeney2021SimonSE}.
Here we aim to evaluate the robustness of \tool against fine-tuning and model pruning.
 

\subsubsection{Model Fine-tuning}

Attackers may fine-tune models with a small amount of data to destroy the authentication scheme of \tool. 
In this section, we take \tool-VGG13 and \tool-ResNet18 trained on CIFAR10 as the baseline models, and then fine-tune the models with new datasets including MNIST, CIFAR100 and GTSRB. 
In each experiment, we fine-tune all the parameters with the Adam optimizer for 50 epochs, and the learning rate is $10^{-4}$. 


The accuracy of the fine-tuned \tool-VGG13 on STL10, GTSRB, and CIFAR100 are 71.14\%, 96.47\%, and 59.31\%, and the corresponding accuracy of the fine-tuned \tool-ResNet on this three dataset are 79.85\%, 98.04\%, and 64.78\% respectively, which means that our fine-tuning is effective.
The results are shown in \tabref{tab:fine_tune}, where the $ACC_{leg}$ and $ACC_{ill}$ are the accuracies on CIFAR10 of the models after fine-tuned. 

Taking \tool-VGG13 as an example, even when the network is fine-tuned with STL10, the $ACC_{ill}$ can only reach 42.57\%, and the results of fine-tuning on other datasets are even worse~(a minimum of 21.99\%). 
This is a failed attack for the attackers, as the network would not leak the correct prediction without mask after fine-tuning process. After fine-tuning for AuthNet-resnet, the model still exhibits a noticeable accuracy difference on legitimate and illegal inputs (6.53\%-30.75\%), which suggests that the authentication mechanism has been retained to a certain extent. For AuthNet-VGG13 on gtsrb and CIFAR100 datasets, the gap between $ACC_leg$ and $ACC_ill$ is small. Therefore, we recommend that for model owners who are allowed to fine-tune or transfer learning the model, they can first fit the model parameters to new task through transfer learning and then embed the authentication logic with our scheme. Therefore, We demonstrate the good robustness of our proposed scheme against the fine-tuning process.

\begin{table}[h]
\scriptsize
\centering
\caption{Robustness of \tool after model fine-tuning. ``Datasets'' indicates the data used for fine-tuning, and ``-'' is the performance of \tool before fine-tuning.}
\label{tab:fine_tune}
\begin{tabular}{cccc}
\toprule
\multicolumn{1}{c}{\textbf{Model}}&
\multicolumn{1}{c}{\textbf{Datasets}}  & 
\multicolumn{1}{c}{\textbf{$ACC_{leg}$}} & 
\multicolumn{1}{c}{\textbf{$ACC_{ill}$}}\\
\midrule

\multirow{4}{*}{VGG13} & {-}& {86.95\%}& {23.03\%}\\
&{STL10} &  {60.87\%}& \textbf{42.57\%}\\
&{GTSRB} &  {22.52\%}& \underline{21.99\%}\\
&{CIFAR100} & {36.17\%}& {38.19\%}\\
\hline
\multirow{4}{*}{ResNet} & {-} & {90.63\%}& {21.70\%}\\
&{STL10} & {79.49\%}& \textbf{48.74\%}\\
&{GTSRB} & {41.12\%}& \underline{22.25\%}\\
&{CIFAR100}& {51.49\%}& {44.96\%}\\
\bottomrule
\end{tabular}
\vspace{-5mm}
\end{table}
\subsubsection{Model Pruning}

\begin{figure}[!t]
\centering
\epsfig{figure=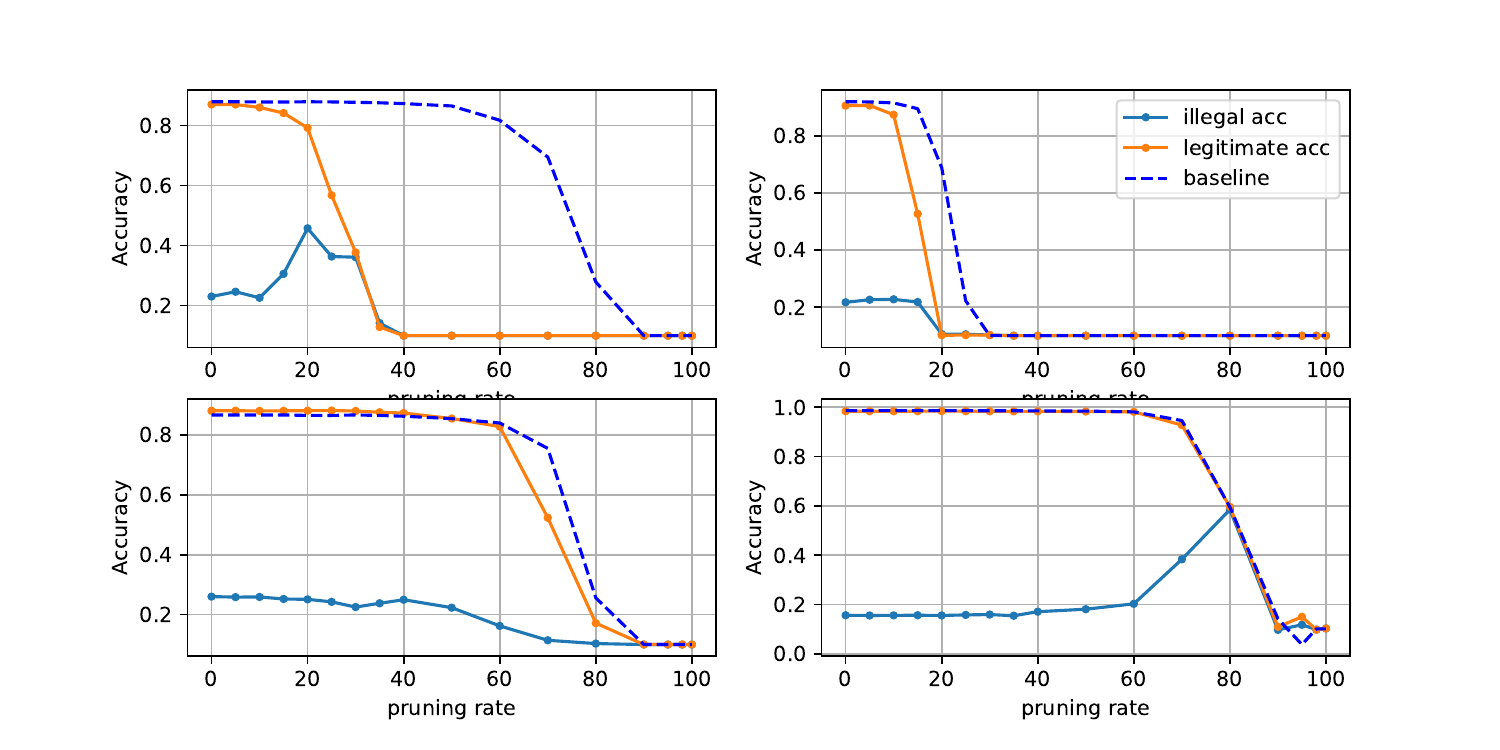, width=0.5\textwidth} 
\caption{Pruning attack on \tool, from left to right and top to bottom, the sub-images represent \tool-VGG13, \tool-ResNet18, \tool-MobileNet, and \tool-LeNet, respectively.}
\label{fig:prune_attack}
\vspace{-8pt}
\end{figure}

Pruning is a common practice in model optimization to compress a neural network by eliminating redundant neurons under the premise of maintaining the accuracy of the original inference task. 
In general, neurons with lower weights or less-connected tend to be the focus of attention during pruning, since they are less capable of conveying information to deeper networks. 
We adapt the commonly used prune technique in \cite{Han2015LearningBW} on four different pre-trained models and their corresponding \tool in \secref{sec:effect}, and the results are shown in \figref{fig:prune_attack}.

In our experiments, the pruning rate varies from 0\% to 100\%.
When the pruning rate is below 40\%, our interval is set to 5\%, and after exceeding 40\%, the pruning rate interval of each set of experiments is 10\%. 
This is because the performance of \tool changes significantly when the pruning rate is lower than 40\%, and we are more concerned about whether the model will leak the correct prediction results to illegal queries during the pruning process of \tool. 
The dashed line in the figure shows the change trend of the classification accuracy of the pre-trained model as the pruning rate keeps increasing. 
For the four pre-trained models, the accuracy starts to decline significantly when the pruning rate exceeds 60\%. This means that the pre-trained model does contain a large number of redundant neurons, in the other words, our theoretical assumption about the redundant neurons in networks is reasonable. 
\tool generally starts to experience a significant decline in accuracy for legitimate users after the pruning rate exceeds 20\% or 30\%. 
This directly means that the proportion of redundant neurons in the model is decreasing, and our scheme does effectively reuse the redundant neurons in the model, which guarantees the authentication function of the model. 
Additionally, the accuracy of illegal queries always stays below 50\% throughout the pruning process, which means that our model also does not leak the correct predicted label during pruning. In summary, our model is robust and secure for the case of pruning.

\subsection{Security Analysis}
\label{sec:security}

We evaluate \tool's security from two aspects, i.e., the confidentiality of \tool mask and the resistency against closed-box attacks (i.e., model extraction and authentication offsetting).

\subsubsection{Mask Inversion Attack}
Here we propose two types of inversion attacks based on different adversary capabilities.

\noindent\textbf{Differential Attack}. Attackers are supposed to have obtained a certain amount of authorized images and their originals. They can compute the differences of images and construct candidate masks. 
Assume $\{\langle img_i, img'_i\rangle \}_{N}$ is the leaked data set, where $img_i$ is the original image before enclosed with \emph{authentication mask}, $img'_i$ is the authorized image accordingly, and $N$ is the number of image pairs. 
Therefore, the mask can be computed as follows.
\begin{equation}
\centering
Mask_D=\frac{1}{N}\sum_{i=1}^{N}(img^{'}_i-img_i)
\label{lossfunction}
\end{equation}

After obtaining $Mask_D$, the attacker superimposes it on input images and queries the network directly with the processed images.
Although the mask can be subtle, it is fixed for a certain \tool. 
Therefore, a considerable number of leaked input images may make the attack feasible. 
To alleviate this security threat, we propose a defense strategy by adding random noise to the input images, i.e., $\epsilon(img')$, where $\epsilon(\cdot)$ is a noising function implemented with the \textit{PIL} package. With this new strategy, the pixel variation of each image is related to the texture of the image itself. 

Tabel~\ref{tab:DA} shows the results for differential attacks on two models, \tool-VGG13 and \tool-AlexNet respectively.
The $ACC_{leg}$ has a slight increase, since the noising function $\epsilon(\cdot)$ chosen here would enhance the texture of the images. 
To a certain extent, \tool can effectively resist the differential attack, dropping by 40\% in the accuracies on different models. 
With the noising function, the resistance against this kind of attack could be more thorough. The total number ($N$) of the collected images would not affect the conclusion within a certain range~($N= 100, 200$ or $500$).

\begin{table}
\footnotesize
\centering
\caption{Differential attack on two models. ``$ACC_{diff}$'' indicates the model accuracy with computed masks. ``N=100'' means the attacker has obtained 100 pairs of original and authorized images.} 
\label{tab:DA}
\begin{tabular}{cccccc}

\toprule

\multirow{2}{*}{\textbf{Model}}&
\multirow{2}{*}{\textbf{Setting}}&
\multirow{2}{*}{\textbf{$ACC_{leg}$}} & 
\multicolumn{3}{c}{\textbf{$ACC_{diff}$}}  \\ \cline{4-6}
& & & \textbf{N=100} & \textbf{N=200} & \textbf{N=500} \\
\midrule

\multirow{2}{*}{VGG13} & {w/o noise} & {86.95\%} & {42.29\%} & \textbf{38.30\%} &{39.62\%}   \\
&{w/ noise}&{87.36\%} & \textbf{28.90\%} & {29.59\%} &{29.62\%}\\
\hline
\multirow{2}{*}{AlexNet} & {w/o noise} & {85.12\%} &{48.28\%} & \textbf{46.76\%} & {49.13\%} \\
&{w/ noise}&{86.46\%} & {42.85\%} & {39.22\%}& \textbf{37.30\%}\\
\bottomrule
\end{tabular}
\end{table}
\vspace{-2pt}
\noindent\textbf{Mask Optimization Attack.} In this scenario, attackers possess a number of original images and their labels. Then they search for an optimal mask, with which the inputs can be correctly classified by the \tool-enhanced model.
Besides, we assume that the attacker has access to \tool, so it can use reverse engineering to optimize a fake key $Mask_{I}$, where a mask and an offset images are trained with \tool parameters frozen. 
The training objective is to improve the \tool accuracy on images embedded with $Mask_{I}$. 
We conduct experiments on \tool-LeNet, \tool-ResNet18, \tool-VGG13 and \tool-ResNet50, using MNIST, CIFAR10, CIFAR10 and CIFAR100, respectively. Additionally, we run 10 epochs for each attack and select the best accuracy. 

\begin{figure}[!h]
\centering
\epsfig{figure=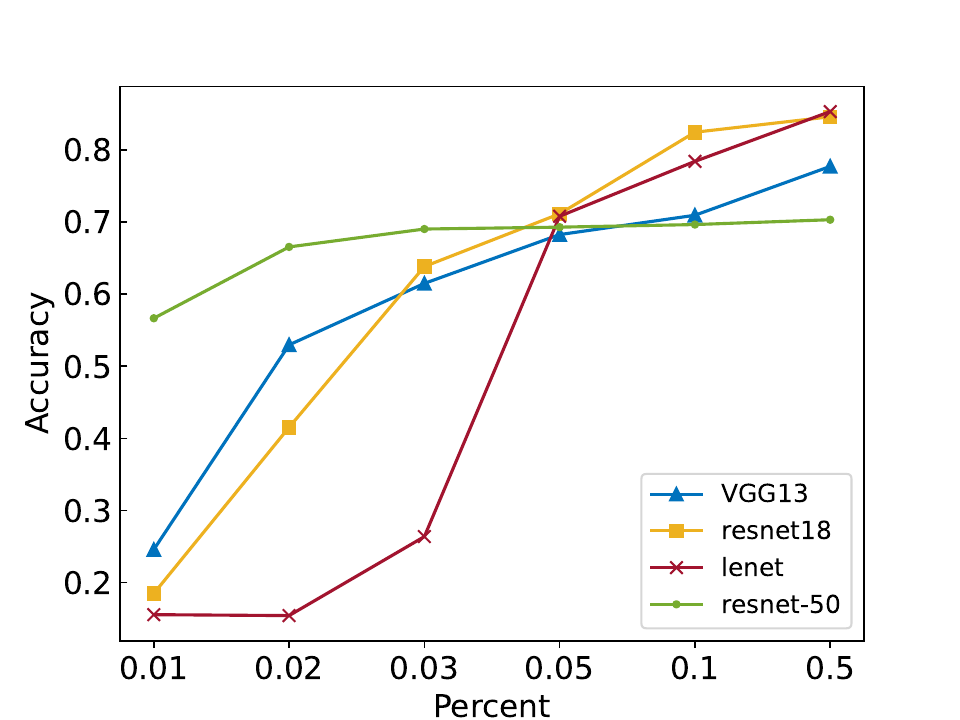, width=0.3\textwidth} 
\caption{The performance of $Mask_{I}$ on \tool}
\label{fig:fake2}
\vspace{-4mm}
\end{figure}

\figref{fig:fake2} presents how the performance of $Mask_{I}$ varies with training set proportions. 
As the proportion of training data obtained by the attacker increases, the performance of $Mask_{I}$ becomes better. 
For \tool-VGG13, using 10\% of training data to reverse $Mask_{I}$, the attacker can gain 70.93\% accuracy, for \tool-ResNet, 5\% of training data is needed to achieve 71.13\% accuracy. 
However, there are still significant gaps to the original model, \i.e., 16.94\% and 20.87\%, respectively. 
To ensure a 10\%-accuracy distance, attackers have to get at least 50\% training data, of which the cost is prohibitive.

In addition, we also find that reverse attacks seems to be more effective for \tool-ResNet50, with large scale parameters, 66.54\% accuracy was achieved with 2\% training data. However, accuracy increases very slowly as the number of training sets increases, 70.31\% accuracy is achieved using 50\% training data, which means that achieving near-clean model performance requires a steep increase in training difficulty. This is related to the generalization problem of large-scale model, a universally applicable key is difficult to fit due to the complex nonlinear transformations of deep learning models.

\subsubsection{Model Extraction}\label{sec:security:modelextraction}
In this scenario, attackers can query \tool's APIs and then train a substitute model with a number of samples and generated soft labels by \tool.

We employed two model extraction approaches, one being a naive model extraction approach, where we do not sample the query dataset and the victim's output is directly used as the teacher for substitute model learning. The other is an existing algorithm named \emph{Knockoff Nets}~\cite{8953839}. 
The general attack framework involves: sampling query samples, obtaining outputs from victim model, and training the substitute model, where the victim model includes \tool-VGG13 and VGG13-baseline, their performance is as depicted in \tabref{tab:effectiveness}. 
The experiment setups for both methods are as follows: the native method utilizes 50,000 CIFAR10 training images as the query dataset, employing Mean Squared Error (MSE) loss function to minimize the soft labels between victim model and substitute model. It is trained for 50 epochs with a learning rate of 0.0001. Knockoff Nets utilizes 100,000 Tiny-ImageNet-200 images as the query dataset, using Cross Entropy as the loss function to reduce the disparity in soft labels between the victim and substitute models. It is trained for 100 epochs with learning rate of 0.01. The model architecture of substitute models chosen by both methods are identical to the victim models. 

\begin{table}[t]
  \centering
  \scriptsize
  \caption{Accuracy of substitutes in model extraction}
    \begin{tabular}{ccc}
    \toprule
     \multicolumn{1}{l}{\textbf{Methods}}&{AuthNet-VGG13} & \multicolumn{1}{l}{VGG13}\\
    \midrule
    Native Method&   {26.48\%}     & {85.57\%} \\
    Knockoff Nets& {26.7\%} &{83.0\%}\\
    \bottomrule
    \end{tabular}%
  \label{tab:subacc}%
  \vspace{-3mm}
\end{table}%
\tabref{tab:subacc} shows the accuracies of trained substitute models by model extraction. 
The result demonstrates a strong protection effect by \tool against query-based model extraction attacks. 
Compared to the notably high 85.57\% and 83.0\% accuracy in model extraction of the original VGG13, \tool-VGG13's substitute models achieve only 26.48\% and 26.7\% accuracy, respectively. 

\setlength{\abovecaptionskip}{0cm}
\begin{table}
  \centering
  \scriptsize
  \caption{Extra layers are added to the models and then fine-tuned to offset the authentication logic in \tool. }
    \begin{tabular}{ccccc}
    \toprule
          & \multicolumn{2}{c}{AlexNet} & \multicolumn{2}{c}{ResNet} \\
\cmidrule{2-5}          & \multicolumn{1}{c}{AuthNet} & \multicolumn{1}{c}{Passport} & \multicolumn{1}{c}{AuthNet} & \multicolumn{1}{c}{Passport} \\
    \midrule
    Ori. Acc.   & 85.58\%     & 89.22\%    & 90.63\%     & 89.67\% \\
    \midrule
    Accuracy   & 12.01\%     & 13.32\%    & 15.95\%     & 59.48\% \\
    \bottomrule
    \end{tabular}%
  \label{tab:extra_layer}%
 \vspace{-5mm}
\end{table}%

\subsubsection{Authentication Offsetting}\label{sec:security:extra_layer} 
Different from fine-tuning the whole model, attackers cannot change the parameters in this scenario, but are able to augment an extra linear layer behind \tool. 
Through meticulous fine-tuning of this supplementary layer, they aim to alleviate the impact of the identity authentication mechanism embedded in the preceding network. 
In particular, the attacker first gains some clean data, adds a fully connected layer to the model, and fine-tunes this layer to correct the classification function, i.e., offsetting identity authentication.


We conduct experiments on ResNet18 and AlexNet, and also provide a comparison with the Passport method. In our adversarial setting, attackers acquire 20\% of CIFAR10 samples and perform fine-tuning for the last additional linear layer. 
The original model's parameters are frozen throughout the entire fine-tuning process. 
After 10,000 rounds of fine-tuning with a learning rate of 0.001 for both ResNet18 
experiments and 0.01 for both AlexNet experiments, the models exhibit the following performance. 
In \tabref{tab:extra_layer}, 
on AlexNet, both approaches exhibit high defensive effects. After fine-tuning, the accuracy rates for AuthNet-AlexNet and Passport-AlexNet are 12.01\% and 13.32\%, respectively. However, on the more substantial and complex model ResNet18, AuthNet's success rate in this attack is significantly lower than that of the Passport method (accuracies on unauthorized data: $11.3\% < 56\%$). It shows that, to a certain extent, AuthNet is more secure than the Passport method in a close-box attack scenario. It is largely because \tool exhibits a stronger confusion effect on refused samples, indicating that our strategy of random output for refused samples is more effective.

\begin{table}[H]
	\scriptsize
	\centering
	\caption{Training and inference costs. ``$T_t(M)$'' denotes the time to train a clean model, ``$T_t(M_{auth})$'' is the time to train a model with authentication logic. \tool's $T_t(M_{auth})$ contains mask inversion time and tail network fine-tuning time. ``$T_i(M)$'' and ``$T_i(M_{auth})$'' denotes the inference time for clean model and model with authentication logic. The inference time of Passport involves passport setting and inference time.}\label{tab:cost}
		\begin{tabular}{ccc}
			\toprule
			\multicolumn{1}{c}{\textbf{Time Cost}}&
			\multicolumn{1}{c}{\textbf{Passport-ResNet18}}&
			\multicolumn{1}{c}{\textbf{\tool-ResNet18}}\\
			\midrule
			
			{$T_t(M_{Auth})$} & {2495.75}&{24.73 + 2067.78} \\
			{$T_t(M)$}&{2168.42}& {1690.59} \\
			{$Cost_t$}&{15.09\%}& {23.77\%} \\
			
			{$T_i(M_{Auth})$} &{0.33+1.80} &{3.20} \\
			{$T_i(M)$}& {1.87}&{3.16} \\
			{$Cost_i$}&{17.65\%}& {1.32\%} \\

			\bottomrule
		\end{tabular}
  \vspace{-2mm}
	\end{table}


\subsection{Cost Analysis} \label{subsec:cost}
To evaluate \tool's practicality, we conduct experiments to measure the cost of model enhancement and inference. 
Additionally, we compare training cost and inference cost of \tool with the Passport method~\cite{NEURIPS2019_75455e06}. 
We test \tool-ResNet18 on dataset CIFAR10, where the hyperparameters of mask inversion and tail network fine-tuning are in \tabref{tab:params}. 
We also conduct the Passport method in the V1 scenario as \cite{NEURIPS2019_75455e06}, since it is aligned with this study.

The results are shown in \tabref{tab:cost}, the Passport layers lead to 10\%-15\% extra running time compared to training a clean model, and about 10\% extra inference time. Nevertheless, for our scheme, since no additional structures are added to the model, embedding logic and inference cost are largely reduced. The extra time cost of training process is 20\%-25\%, and it makes subtle extra cost for inference (only 1.32\%).

\section{Discussion}
\label{sec:disc}
\noindent\textbf{Generalizability.}
We introduce the \tool method and experiments primarily based on computer vision tasks. 
The identity information can be easily embedded into images with minimal perturbations, and then \tool is trained to extract this information and recognize the identity. 
To adapt for other tasks like natural language processing, \tool should be equipped with a new method to encode identity information in tabular or textual input. 
Prior studies~\cite{Zhang2019AdversarialAO,Morris2020TextAttackAF,Cartella2021AdversarialAF,Ebrahimi2017HotFlipWA} propose several methods of evolving words in text or text perturbation to construct adversarial examples or inject backdoors. 
Considering extending \tool to text classification or generative tasks, we can use such an atom-level perturbation for text as secret key embedding process, and select authentication bits on word vector, where the perturbation embedded in the original text would lead to high activation values, then fine-tune the tail network, which is responsible for verifying the validity of the input.

\noindent\textbf{Enhancements for \tool.}
It has a high potential to be enhanced against more advanced attacks. For example, if one attacker employs the brute-force approach in a clear-box setting, identifying the layer containing authentication bits and fine-tuning one offset layer before it. 
The authentication bits can be dispersed into multiple layers and the tail network is after the last layer of authentication bits. We conduct a group of experiments in VGG13 on CIFAR10 dataset to distribute 150 authentication bits across 3 layers with  $\gamma$=30. The initial accuracy for legitimate inputs is 83.16\%, while for illegal inputs, it is 39.09\%. 
Authentication offset attacks with fine-tuning 10\%, 20\% and 50\% of training data can achieve 29.4\%, 40.2\% and 72.06\% accuracy, respectively. 
Even worse, assuming the number of candidate layers is $n$, the complexity of such attacks is raised from $O(n)$ times of fine tuning to $O(2^n)$ if the adversary has no idea about the number of layers with authentication bits.
\section{Related Work}


\noindent{\bf Model Stealing Attacks.} Deep learning models are vulnerable to stealing attacks upon deployed as described below.

Query-based model extraction attacks~\cite{8953839,8489592,251526} query the target model with samples to obtain soft output labels, and a substitute model is then trained based on input-output pairs to extract model functionality. \citet{8953839} propose a scheme in the absence of additional model knowledge to train a substitute model only by querying the black-box model, and the functionality of model can be stolen within 60,000 queries. \citet{8489592} query images in the non-problem domain to train substitute model. The feasibility of model extraction attack is demonstrated in more realistic MLaaS scenarios (e.g. Microsoft Azure Emotion API). \citet{251526} conduct experiments in the ReLU-based fully connected network, and enable a high fidelity recovery of model parameters and functionality by improving search strategy. 
Side-channel attacks \cite{Wei2020LeakyDS,Yu2020DeepEMDN,Wei2018IKW,Batina2018CSINN,Hua2018ReverseEC,Duddu2018StealingNN} is also a serious threat. These side-channel attacks are designed to either capture model structure~\cite{Wei2020LeakyDS,Yu2020DeepEMDN}, or disclose training data privacy~\cite{Wei2018IKW,Duddu2018StealingNN}, or explore the theft of both model parameter and structure~\cite{Batina2018CSINN,Hua2018ReverseEC}. 
\citet{Wei2020LeakyDS} attack the structure of deep learning models in the shared GPU scenario, where model structure and hyperparameters of VGG16 are recovered with 95\% and 82.8\% accuracy. \citet{Yu2020DeepEMDN} use electromagnetic side-channel attacks to steal structure and hyperparameter, then adopt black-box model extraction attack, achieving 96\% performance of original model. \citet{Wei2018IKW} conduct experiment on a convolutional neural network accelerator based on FPGA, and the input image is recovered from the collected power trajectory without knowing the detailed parameters. \citet{Duddu2018StealingNN} uses timing channel to infer hyperparameter of convolutional neural network, which can be used in membership inference attack leading to the leakage of  training data. \citet{Batina2018CSINN} propose a differential power analysis on an ARM CORTEX-M3 microcontoller, allowing attackers to obtain parameter, hyperparameter and structure of model. \citet{Hua2018ReverseEC} extract weight information by constructing samples to observe the output through memory and timing side-channel, which is robust to dynamic zero pruning model transformation and data encryption protection.
\textit{Our solution~\tool can well mitigate the above attacks. On one hand, query-based attacks need send a large number of requests to the model, which can be easily detected and prevented. On the other hand, although attackers can employ side-channel techniques to recover a substitute model, the authentication logic inside the model can still stop attackers to use.}

\vspace{3pt}
\noindent{\bf Model Protection.} To protect intellectual property of deep learning models, there are currently passive and active defenses available in practice~\cite{8630791}. \citet{8630791} firstly propose a scheme, binding model availability to authentication results of users. A mask was given to the legitimate user to ensure the model performance by adding the mask to each image. 
\citet{Tang2020DeepSN} embed 0-1 verification code patch on images in training process of student model in model distillation scenario. When distributing student model, only the correct captcha images can ensure that the model is trained normally.
\citet{NEURIPS2019_75455e06} distribute passport to legitimate users, the passport validation layer in the model will map the layer input linearly, and the model accuracy can only be guaranteed by using the correct passport.
\citet{10.5555/3495724.3497620} place a passport validation logic in the parameters of normalization layer, which does not change model structure when reasoning about the validation. Users who do not provide the correct passport will not be able to obtain high accuracy output of the model.

However, it is witnessed that these protection solutions are vulnerable to ambiguity attacks \cite{10204223,Li2006ZeroknowledgeWD,Sencar2007CombattingAA,Loukhaoukha2017AmbiguityAO}. 
\citet{10204223} is an efficient ambiguity attack on passport-based ownership verification scheme~\cite{NEURIPS2019_75455e06}, where they forge the passport by replacing the passport layer with a fully connected layer to training a new scale factor and bias factor to construct forgery passport. \citet{Li2006ZeroknowledgeWD} propose a zero knowledge watermark detector, which allows model owner completes ownership verification without disclosing verification information to resist ambiguity attack.  \citet{Sencar2007CombattingAA} propose a multiple watermark embedding scheme, which combines a series of selecting detection and unidirectional function in watermark embedding process, the verifier should provide not only secret information but also correct selection strategy in verification process. \citet{Loukhaoukha2017AmbiguityAO} propose an ambiguity attack to generate a robust blind image watermark based on diffuse discrete wavelet transform and singular value decomposition.
\textit{Apart from prior studies, we pick authentication bits on the internal gate layer and add the authentication function on them. It is more flexible and transparent since it is not limited to specific scenarios such as distillation learning and does not modify the model structure.} 

\section{Conclusion}
\label{sec:conclusion}
We propose a novel authentication scheme \tool for neural networks that can autonomously authenticate users and prevent model misuse. 
Differently from traditional methods, \tool's authentication ability is integrated inside the model by reconstructing neural networks with three building blocks: the \emph{head model} is to extract authentication information, which is then encoded in the \emph{gate layer}, and the \emph{tail model} can recognize the identity and make prediction accordingly. 
Through experiments, \tool shows superior performance in user authentication, robustness from model transformations, and resilience to adaptive attacks. 

\bibliographystyle{ACM-Reference-Format}
\bibliography{reference}
\section*{Appendix}\label{sec:appendix}

\titleformat{\subsection}
{\normalfont\large\bfseries}{\Alph{subsection}. }{0em}{}

\subsection{Demonstration for AuthNet's Principles }\label{sec:appendix:principle}
\begin{figure}[!h]
\centering
\epsfig{figure=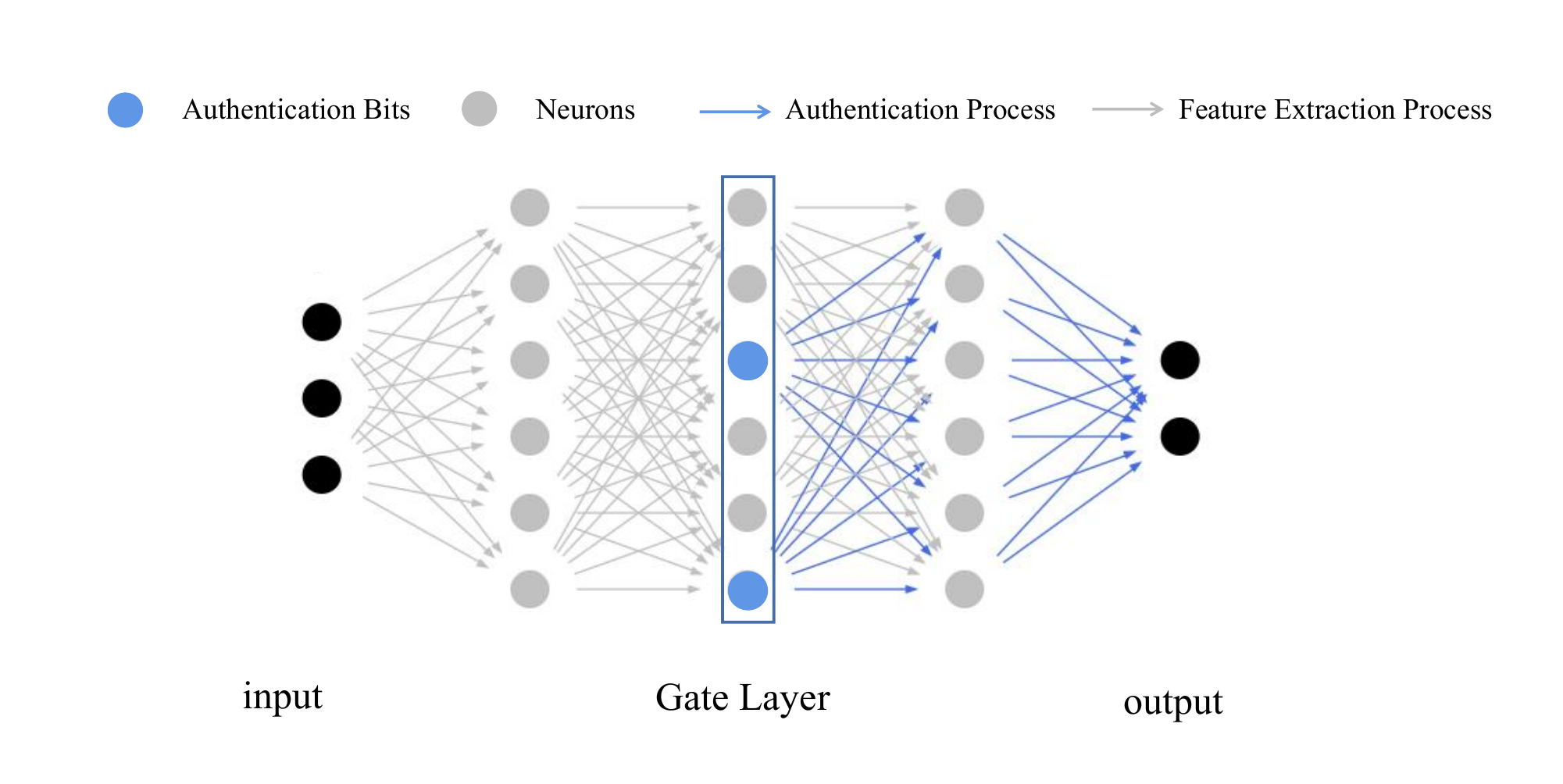, width=0.5\textwidth} 
\caption{Activation values on authentication bits in \tool affect the final outputs.}
\label{fig:AB}
\end{figure}

We adopt a classical model Multilayer Perceptron to illustrate the specific details of the gate layer and authentication bits. As shown in \figref{fig:AB}, each circle is a neuron that represents a convolution operation, receiving signals from different channels and outputs the results of the operation, where we use blue for authentication bits and gray for other neurons. 
After the inference of head model, the non-authentication neurons contain texture information, while the authentication bits have both authentication and texture information. 
The differential performance of authentication bits of legitimate and illegal inputs will affect the tail model neurons diffusively, which finally causes different predictive output to achieve the purpose of authentication.

\subsection{Hyperparameters of Models}\label{sec:appendix:hyperparameter}


\begin{table}[H]
\footnotesize
\caption{The sequence of layers in VGG13 (dropout $p=0.5$, batch size=256)}
\label{tab:VGG13}
\begin{tabular}{cccc}
\toprule
\multicolumn{1}{c}{\textbf{Seq.}} & \multicolumn{1}{c}{\textbf{Layer Type}} & \multicolumn{1}{c}{\textbf{Kernal Size}} &\multicolumn{1}{c}{\textbf{Output Size}} \\ 

\midrule

{1} & {Conv} & {$64\times3\times3\times3$} &{$256\times64\times32\times32$} \\
{2} & {Conv} & {$64\times3\times3\times64$} & {$256\times64\times32\times32$}\\
{3} & {AVG Pool} & {$2\times2$} & {$256\times64\times16\times16$}\\ 
{4} & {Conv} & {$128\times3\times3\times64$} & {$256\times128\times16\times16$}\\
{5} & {Conv} & {$128\times3\times3\times128$} & {$256\times128\times16\times16$}\\
{6} & {AVG Pool} & {$2\times2$} & {$256\times128\times8\times8$}\\ 
{7} & {Conv} & {$256\times3\times3\times128$} & {$256\times256\times8\times8$}\\
{8} & {Conv} & {$256\times3\times3\times256$} & {$256\times256\times8\times8$}\\
{9} & {AVG Pool} & {$2\times2$} & {$256\times256\times4\times4$}\\
{10} & {Conv} & {$512\times3\times3\times256$} & {$256\times512\times4\times4$}\\
{11} & {Conv} & {$512\times3\times3\times512$} & {$256\times512\times4\times4$}\\
{12} & {AVG Pool} & {$2\times2$} & {$256\times512\times2\times2$}\\
{13} & {Conv} & {$512\times3\times3\times512$} & {$256\times512\times2\times2$}\\
{14} & {Conv} & {$512\times3\times3\times512$} & {$256\times512\times2\times2$}\\
{15} & {AVG Pool} & {$2\times2$} & {$256\times512\times1\times1$}\\
{16} & {F.C.} & {$512\times100$} & {$256\times100$}\\
{17} & {F.C.} & {$100\times10$} & {$256\times10$}\\
\bottomrule

\end{tabular}
\end{table}


\begin{figure*}[btp]
	\centering
	\begin{subfigure}{0.325\textwidth}
		\centering
		\includegraphics[width=0.9\linewidth]{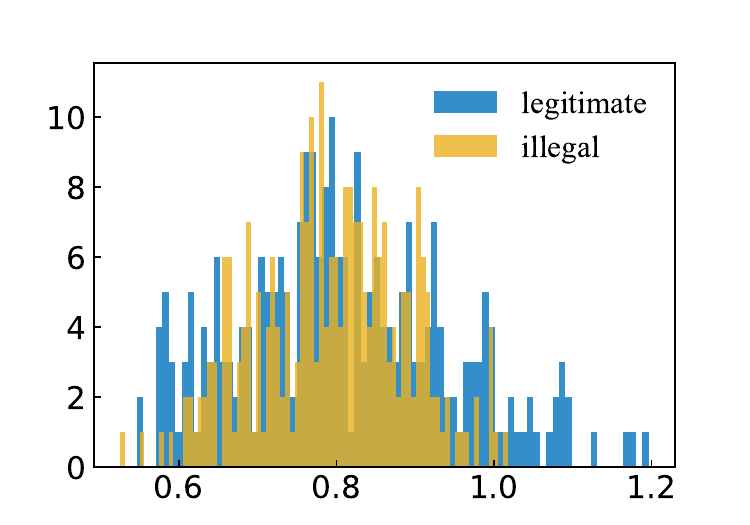}
		\caption{Histogram differences of inter activation}
		\label{hist}
	\end{subfigure}
	\centering
	\begin{subfigure}{0.325\textwidth}
		\centering
		\includegraphics[width=0.9\linewidth]{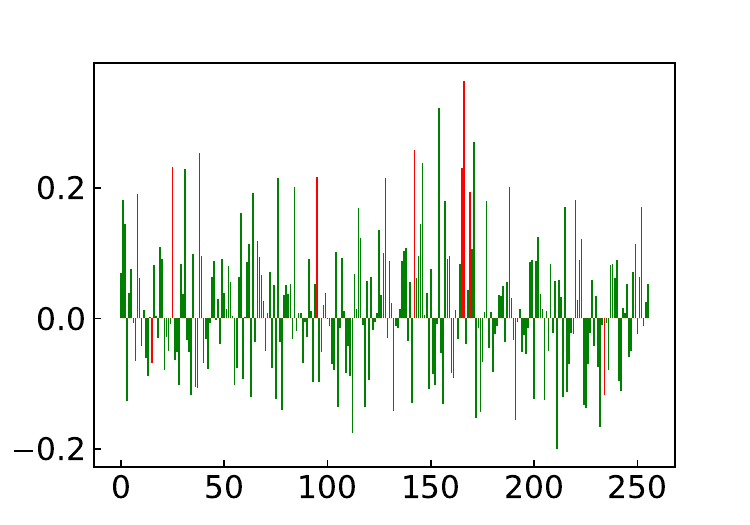}
		\caption{The variation of activation in gate later}
		\label{difference}
	\end{subfigure}
	\centering
	\begin{subfigure}{0.325\textwidth}
		\centering
		\includegraphics[width=0.9\linewidth]{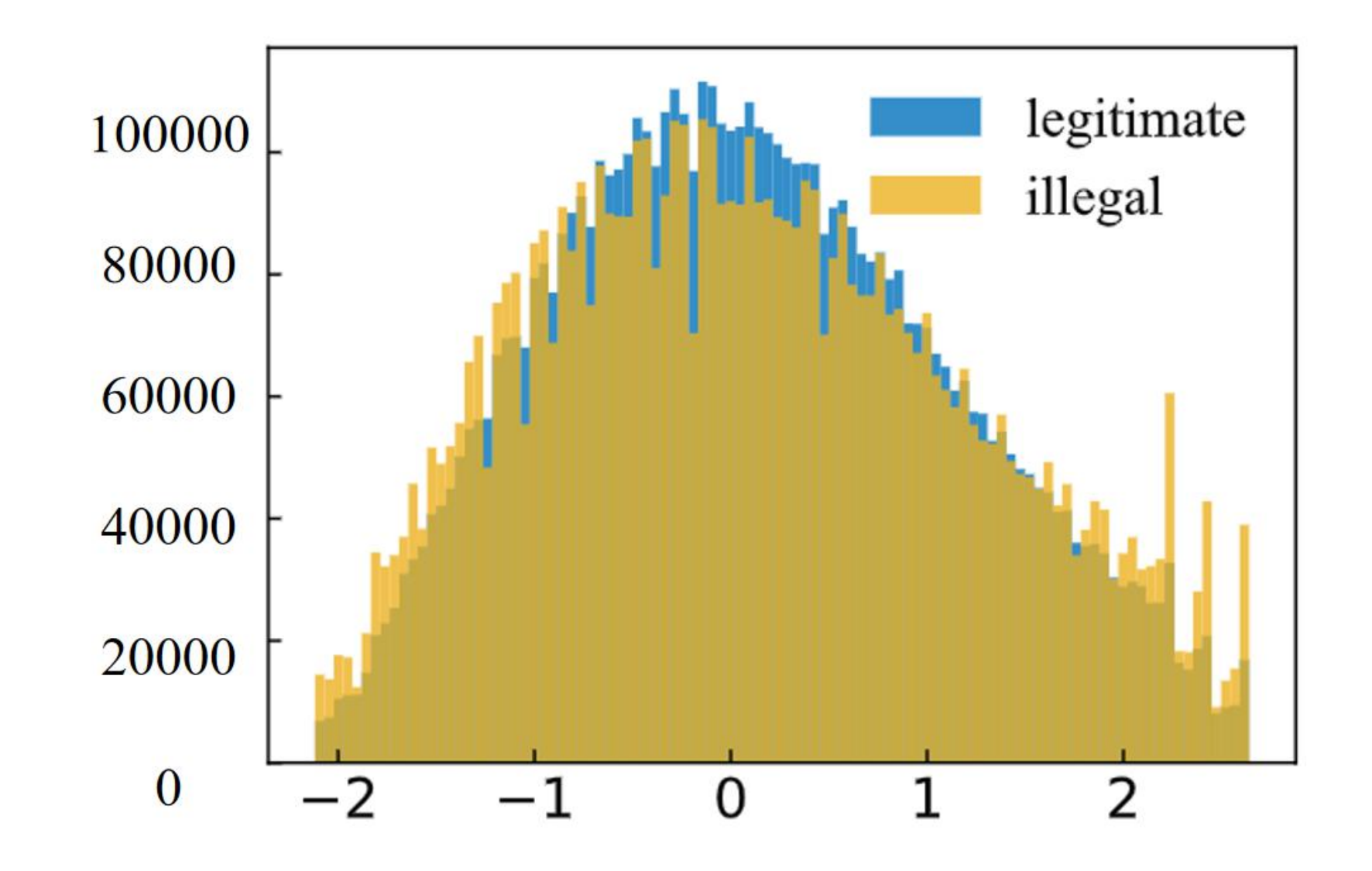}
		\caption{The distribution difference of images}
		\label{image_difference}
	\end{subfigure}
	\\
	\caption{The differences between legitimate and illegal images inferred by \tool-VGG13. The x axis in histogram (a) is the inter activation values of neurons in gate layer. (b) is the average activation variation of different neurons with 10,000 pairs of legitimate and illegal images, where the x axis represents all 256 neurons in gate layer of \tool-VGG13. (c) is the distribution difference of two datasets consisted of 2,000 legitimate images and 2,000 illegal images, respectively.}
	\label{compare_fig}
\end{figure*}

\subsection{Transparency of \tool}\label{sec:appendix:transparency}



We show the histogram of activation values on gate layer in \figref{hist}. Totally, the distribution of activation values on gate layer remains consistent, the secret key alters the activation of interneurons of gate layer by an average of 2.24\% on total neurons, 24.24\% on authentication bits.
It indicates that \tool achieves authentication logic by identifying the activation degree of authentication bits.
As shown in \figref{difference}, we compare the activation value of each neuron at the gate layer before and after the secret key embedding. The experiment is conducted on \tool-VGG13 and CIFAR10, where 10,000 legitimate images and 10,000 illegal images are used. The red bar indicates the location of authentication bits.
The results show that the secret key generally increases the activation value of authentication bits, for other neurons, however, the change of direction is unpredictable and on the same order of magnitude as authentication bits. 
It concludes that authentication bits are well concealed, since an attacker cannot find the location by comparing the intermediate features of gate layer.
The other aspect of concealment is the slight difference between legitimate and illegal inputs. As shown in \figref{image_difference}, we present the distribution of legitimate inputs and illegal inputs adopting 2,000 CIFAR10 images as illegal inputs and 2,000 corresponding processed images with secret key. Compared with illegal images, the mean and variance of the data distribution of legitimate images changed within 15\%~(7.55\% and 14.75\%, respectively), indicating a high similarity of these two distributions. 

\begin{figure}[H]
\centering
\includegraphics[width=0.5\textwidth]{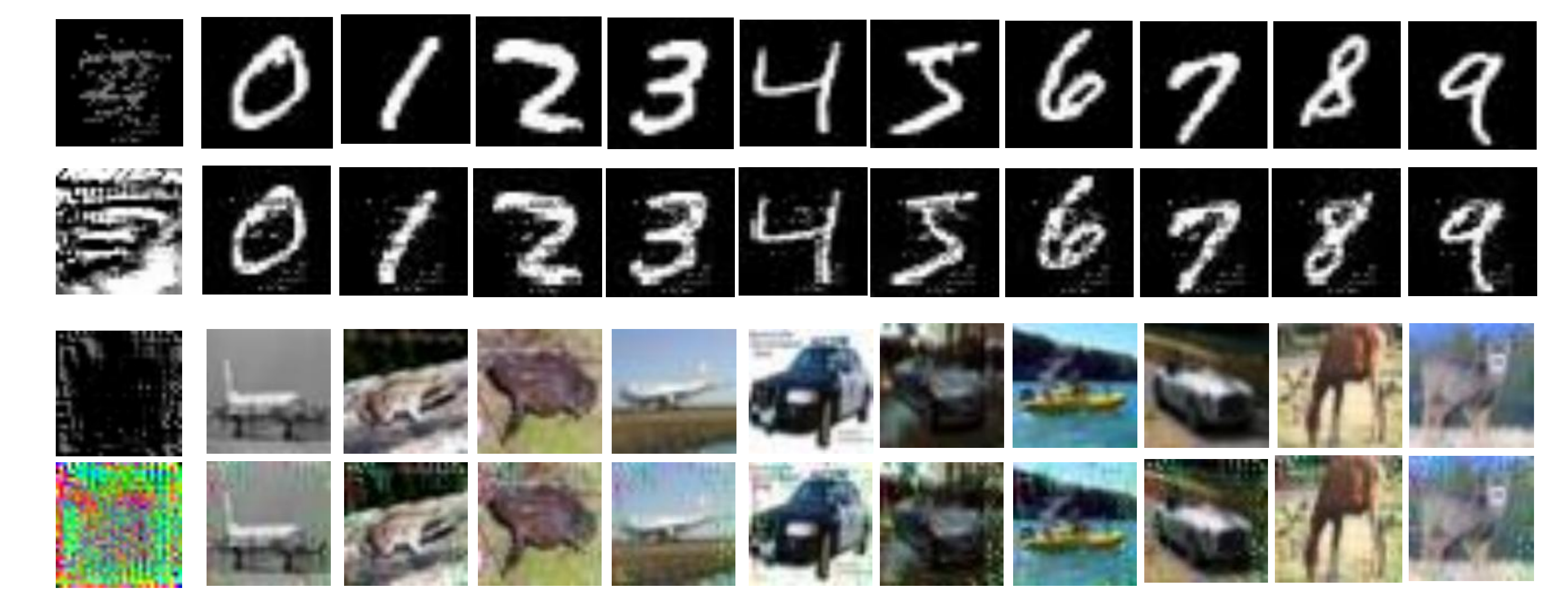}
\caption{Identity mask, original images and processed images with masks}
\label{fig:visualizations}
\end{figure}

In addition, we also present the visualization of legitimate and illegal images in \figref{fig:visualizations}, showing the invisibility of our secret key. 
The first column of \figref{fig:visualizations} is the identity \textit{key} of \tool. More specifically, the gray-scale \textit{mask} controls the location, and the RGB \textit{offset} contains the numerical information of the texture.
The four rows from top to bottom correspond to the original images from MNIST, the preprocessed images from \tool-LeNet, the original images from the CIFAR10 dataset, and the preprocessed images from the \tool-VGG13. 
From the figure, it shows that the difference between the masked image and the original image is trivial and may not raise the awareness of attackers.




\setlength{\abovecaptionskip}{0cm} 

\subsection{Supplementary Details of Theoretical Analysis}\label{sec:appendix:theory}

    

\begin{figure*}[!h]
\centering
\includegraphics[width=0.7\textwidth]{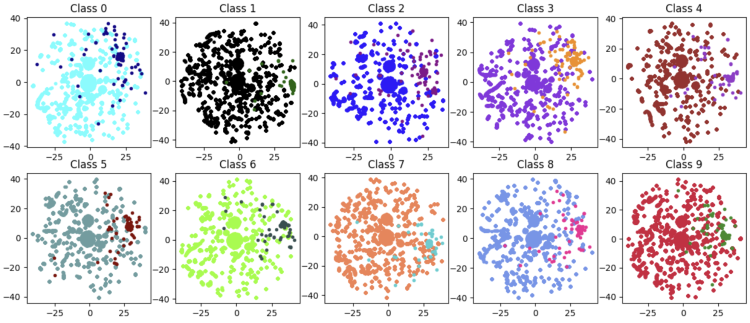}
\caption{t-SNE dimensionality reduction plots representing the refuse domain and authentication domain}
\label{fig:t-SNE}
\end{figure*}

\begin{figure*}
  \centering
  \begin{subfigure}{\textwidth}
    \centering
    \includegraphics[width=0.7\linewidth]{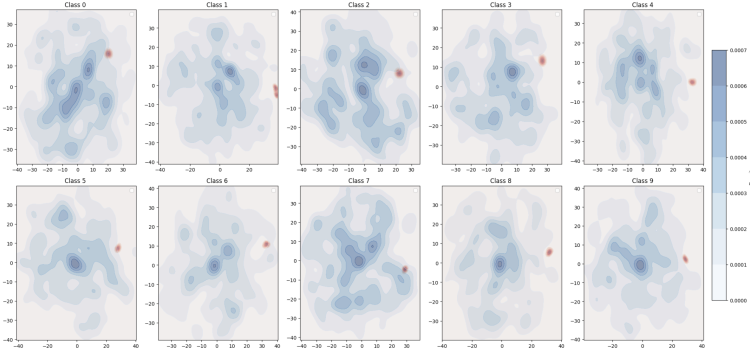}
    \caption{The graphical representation of the KDE analysis performed on the t-SNE dimensionally reduced sample domain. The blue region signifies the refuse domain, while the red region represents the authentication domain. The shading in the image indicates the density of the sample regions. All samples adhere to a consistent evaluation.}
    \label{fig:kde-raw}
  \end{subfigure}
  
  \begin{subfigure}{\textwidth}
    \centering
    \includegraphics[width=0.7\linewidth]{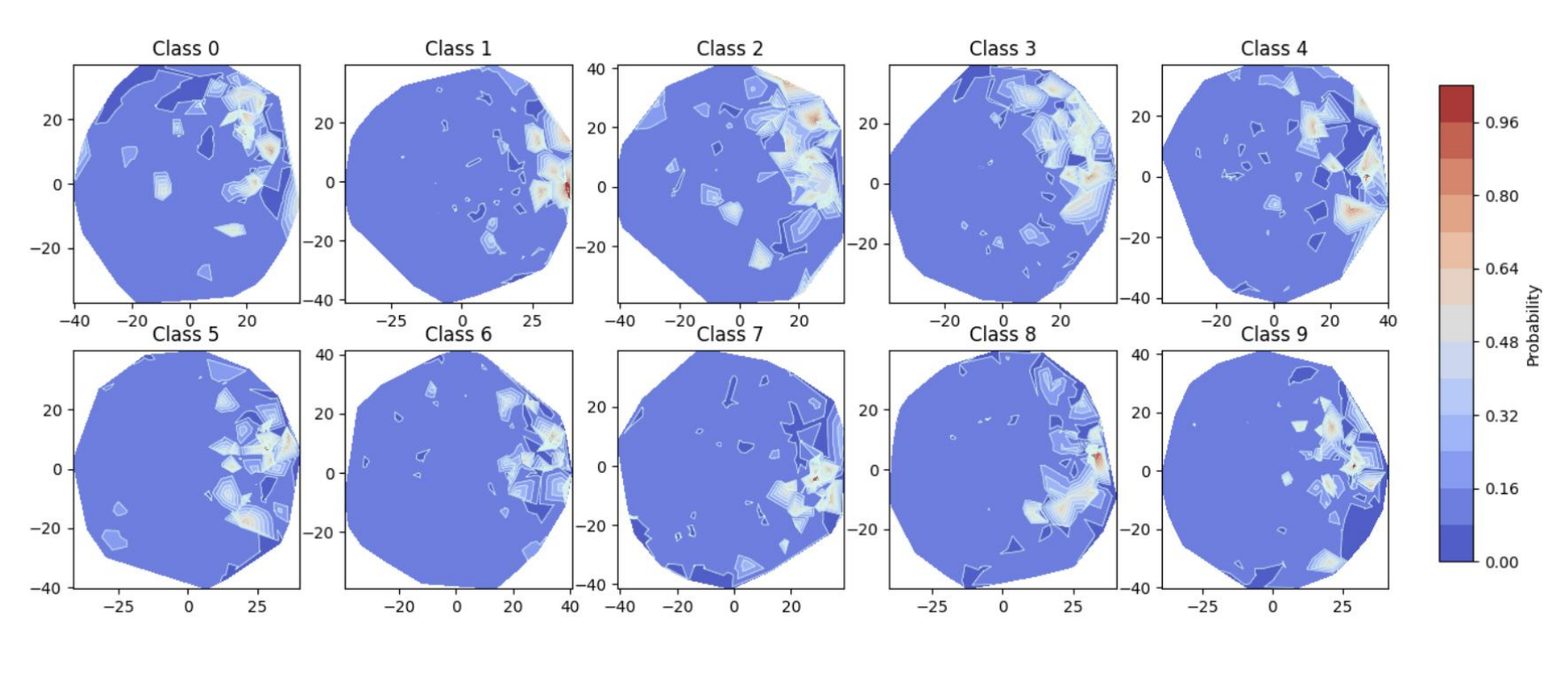}
    \caption{The contour plot represents the accuracy of all samples. Regions depicted in red indicate higher sample accuracy, while cooler tones indicate lower accuracy.}
    \label{fig:kde-acc}
  \end{subfigure}
  
  \caption{The density distribution plot on the dimensionally reduced samples along with the contour plot representing the accuracy on the same reduced samples.}
  \label{fig:kde-compelete}
\end{figure*}

\end{document}